\documentstyle[multicol,pra,aps,epsfig]{revtex}
\begin{document}
\draft 

\title{Decay of an excited atom near an absorbing 
microsphere}

\author{Ho Trung Dung\cite{byline}, 
Ludwig Kn\"{o}ll, and Dirk-Gunnar Welsch}
\address{
Theoretisch-Physikalisches Institut, 
Friedrich-Schiller-Universit\"{a}t Jena, 
Max-Wien-Platz 1, 07743 Jena, Germany}

\date{December 19, 2000}
\maketitle

\begin{abstract}
Spontaneous decay of an excited atom near a dispersing and
absorbing microsphere of given complex permittivity that
satisfies the Kramers-Kronig relations is studied, with special
emphasis on a Drude-Lorentz permittivity. Both the whispering
gallery field resonances below the band gap (for a dielectric
sphere) and the surface-guided field resonances inside the gap
(for a dielectric or a metallic sphere) are considered. Since the
decay rate mimics the spectral density of the sphere-assisted 
ground-state fluctuation of the radiation field, the
strengths and widths of the field resonances essentially
determine the feasible enhancement of spontaneous decay. 
In particular, strong enhancement can be observed for transition
frequencies within the interval in which the surface-guided
field resonances strongly overlap. 
When material absorption becomes significant,
then the highly structured emission pattern that can be
observed when radiative losses dominate reduces to that
of a strongly absorbing mirror.
Accordingly, nonradiative decay becomes dominant.
In particular, if the distance between the atom and
the surface of the microsphere is small enough,
the decay becomes purely nonradiative.   
\end{abstract}

\pacs{PACS numbers: 42.50.Ct, 42.60.Da, 42.50.Lc}

\begin{multicols}{2}

\section{Introduction}
\label{sec1}

Light propagating in dielectric spheres can be trapped by repeated
total internal reflections. When the round-trip optical paths
fit integer numbers of the wavelength, whispering gallery (WG)
waves are formed, which combine extreme photonic confinement with 
very high quality factors. The frequencies and linewidths of 
WG waves are highly sensitive to strain, temperature, 
and other parameters of the surrounding environment in general.
The unique properties of WG waves are crucial to cavity-QED
and various optoelectronics applications~\cite{1}.
WG waves with $Q$-values larger than $10^9$ have been
reported for fused-silica microspheres~\cite{2,3,4} and
liquid hydrogen droplets~\cite{5}, and the ultimate 
level determined by intrinsic material absorption has been
achieved~\cite{3}.

Since the spontaneous decay of an excited atom depends on both
the atom and the ambient medium \cite{6}, it can be expected
that if an atom is near a dielectric microsphere, its spontaneous
decay sensitively responds to the sphere-assisted electromagnetic
field structure. In the case of an atom in free space,
the well-known continuum of free-space density of modes
of radiation is available for an emitted photon. By redistributing
the density of possible field excitations in the presence of
dielectric bodies, the radiative properties of excited atoms can be
controlled. Experimental observations of lifetime modifications of 
ions or dye molecules embedded in microspheres or droplets \cite{9}, 
cavity QED effects in the coupling of a dilute cesium vapor
to the external evanescent field of a WG mode in a fused silica 
microsphere \cite{10}, and detection of individual spatially
constrained and oriented molecules on the surfaces of glycerol
microdroplets~\cite{11} have been reported.
In Refs.~\cite{7,8}, the decay rate of a two-level atom located
in the vicinity of a dielectric sphere was calculated and
enhancement factors of hundreds were predicted. 
The calculations are based on the assumption that there
is no material absorption. In practice however, there exists always
some material absorption that, even when it is small, can still
affect the system in a substantial way. Knowledge
of both the radiative losses and the losses due
to material absorption is crucial, e.g., in designing
ultralow threshold microsphere lasers using cooled atoms or
single quantum dots \cite{13}.

In the present paper we study the problem of spontaneous decay
of an excited atom near an absorbing microsphere of
given complex permittivity that satisfies the
Kramers-Kronig relations, applying the formalism
developed in Refs.~\cite{12,19a}. This enables us
to take into account both dispersion and absorption in
a consistent way, avoiding any restrictive conditions
with respect to the frequency domain. In the numerical
calculations we assume that the permittivity of the microsphere
depends on the frequency according to the Drude-Lorentz model,
which covers dielectric and metallic matter. 
  
The paper is organized as follows. In Sec.~\ref{sec2}  
the radiation field resonances associated with a microsphere
are examined. Equations for determining 
the positions and widths of both the WG
field resonances below a band gap and the
surface-guided (SG) field resonances inside a band gap
are given.
The corresponding quality factors are calculated
and the effects of dispersion, absorption, and 
cavity size are studied.  
Further, the spatial distribution of the cavity-assisted
ground-state fluctuation of the radiation field is studied.
In particular, the wings outside the microsphere of the
fluctuation of the WG and SG waves are just responsible for
exciting such waves in the spontaneous decay of an
excited atom that is situated near the surface of the sphere. 

In Sec.~\ref{sec3} the rate of spontaneous decay, the
line shift, and the (spatially resolved) intensity of the
emitted light are calculated as functions of the atomic
transition

\begin{figure}[!t!]
\noindent
\begin{center}
\epsfig{figure=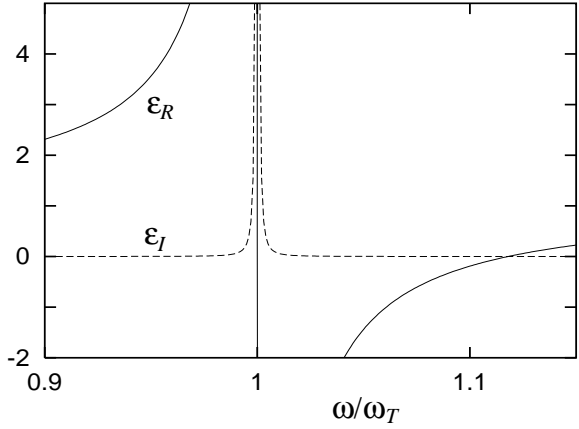,width=1.\linewidth}
\end{center}
\caption{ 
Real and imaginary parts of the permittivity of a 
Drude-Lorentz-type dielectric, Eq.~(\protect\ref{e10}), for 
\mbox{$\omega_P$ $\!=$ $\!0.5\,\omega_T$} and
\mbox{$\gamma$ $\!=$ $\!10^{-4}\omega_T$}.
The band gap covers the interval from $\omega_T$ to 
\mbox{$\omega_L$ $\!\simeq$ $\!1.12\,\omega_T$}.
}
\label{f1a}
\end{figure}
\noindent
frequency, with special emphasis on the influence
of the orientation of the transition dipole moment, the
distance between the atom and the microsphere, and the 
dispersion and absorption outside and inside a band gap. In 
this context, the ratio of radiative decay to
nonradiative decay due to material absorption is analyzed. 

In Sec.~\ref{sec4} the results derived for the
general case of dielectric sphere are applied to a
metallic microsphere by appropriately specifying the
matter parameters. Typical features are briefly discussed. 
Finally, some concluding remarks are given in Sec.~\ref{sec5}. 
 

\section{Radiation field resonances}
\label{sec2}

Let us consider a
dielectric sphere of radius $R$ 
whose complex permittivity $\epsilon(\omega)$ $\!=$
$\!\epsilon_R(\omega)$ $\!+$ $\!i\epsilon_I(\omega)$ reads,
according to a single-resonance Drude-Lorentz model,
\begin{equation}
\label{e10}
        \epsilon(\omega) = 1 + 
        {\omega_P^2 \over 
        \omega_T^2 - \omega^2 - i\omega \gamma}\,,
\end{equation}
where $\omega_P$ corresponds to the coupling constant, 
and $\omega_T$ and $\gamma$ are respectively the medium oscillation
frequency and the linewidth. An example of the dependence 
on frequency of the permittivity is shown in Fig.~\ref{f1a}.
Note that $\epsilon(\omega)$ satisfies the Kramers-Kronig
relations. {F}rom the permittivity, the complex refractive index can
be obtained according to the relations 
\begin{eqnarray}
\label{e1}
    &n(\omega)=\sqrt{\epsilon(\omega)}=n_R(\omega)
     + in_I(\omega),
\\ &\displaystyle\label{e1a}
     n_{R(I)}(\omega) = 
     \sqrt{{1\over2}\left[
     \sqrt{\epsilon_R^2(\omega)+\epsilon_I^2(\omega)}
     +(-)\, \epsilon_R (\omega)\right]} \ .
\end{eqnarray}
The dielectric features a band gap between the transverse 
frequency $\omega_T$ and the longitudinal frequency
\mbox{$\omega_L$ $\!=$ $\!\sqrt{\omega_T^2+\omega_P^2}$}.
Far from the medium resonances,
we typically observe that
\begin{equation}
\label{e1b}
\epsilon_I(\omega) \ll |\epsilon_R(\omega)|.
\end{equation}
For $\omega$ $\!<$ $\!\omega_T$, i.e., outside the
band gap, we have
\begin{eqnarray}
\label{e10a}
     &\epsilon_R(\omega)>1,
\label{e10a1}\\&\displaystyle
     n_R(\omega)\simeq\sqrt{\epsilon_R(\omega)}\gg
     n_I(\omega)\simeq {\epsilon_I(\omega)\over
     2\sqrt{\epsilon_R(\omega)}}\,, 
\end{eqnarray}
and for $\omega_T$ $\!<$ $\!\omega$ $\!<\omega_L$,
i.e., inside the band gap, 
\begin{eqnarray}
\label{e10b}
     &\epsilon_R(\omega)<0,
\\&\displaystyle
     n_R(\omega)\simeq {\epsilon_I(\omega)\over
     2\sqrt{|\epsilon_R(\omega)|}}\ll
     n_I(\omega)\simeq \sqrt{|\epsilon_R(\omega)|}\,.  
\label{e10b1}
\end{eqnarray}
Note that $\epsilon_I(\omega)$ $\!>$ $\!0$ in any case. 

Using the source-quantity representation of the quantized
electromagnetic field \cite{19a,13a}, all the information about the 
influence of the sphere on the spontaneous
decay of an atom is contained in the Green tensor
of the classical, phenomenological Maxwell equations,
where dispersion and absorption are fully taken into
account \cite{12,19a}. In particular, the sphere-assisted
(transverse) field resonances that can give rise to an enhancement of
the spontaneous decay are determined by the poles
of the transverse part of the Green tensor.  

{F}rom the explicit form of the Green tensor as given in
Appendix A, it follows, on setting the denominators in
Eqs.~(\ref{A6}) [or (\ref{A7a})] and (\ref{A7}) [or (\ref{A7b})]
equal to zero, that the resonances are the complex roots 
\begin{equation}
\label{e2}
     \omega = \Omega - i \delta
\end{equation}
of the characteristic equations
\begin{equation}
\label{e3}
     M(\omega)= 
     {H'_{l+1/2}(k_1R) \over H_{l+1/2}(k_1R)}
     -\sqrt{\epsilon(\omega)}\,
     {J'_{l+1/2}(k_2R) \over J_{l+1/2}(k_2R)} = 0 
\end{equation}
for TE waves, and
\begin{eqnarray}
\label{e4}
\lefteqn{
     M(\omega)= 
     {H'_{l+1/2}(k_1R) \over H_{l+1/2}(k_1R)}
}
\nonumber\\&&\hspace{2ex}
      -{1\over \sqrt{\epsilon(\omega)}}
     {J'_{l+1/2}(k_2R) \over J_{l+1/2}(k_2R)} 
     + {1\over 2k_1R}\left[1-{1\over\epsilon(\omega)} \right] 
     = 0 
\end{eqnarray}
for TM waves, where $k_1$ $\!=$ $\!\omega/c$, 
$k_2$ $\!=$ $\!\sqrt{\epsilon(\omega)}\omega/c$, and $R$ is
the microsphere radius 
[$J_\nu(z)$ - Bessel function; $H_\nu(z)$ - Hankel function]. 
In Eq.~(\ref{e2}), $\Omega$ and $\delta$, respectively,
are the position and the HWHM of a resonance line.
Equations (\ref{e3}) and (\ref{e4}) are in agreement with
the results of classical Lorenz-Mie scattering theory \cite{1}.
Similar equations with real permittivity have also been
derived quantum mechanically by expanding the field operators 
in spherical wave functions \cite{13b}. Note that
these results cannot be extended to absorbing media
by simply replacing the real permittivity by a complex one, 
but requires a more refined approach to the electromagnetic 
field quantization in absorbing dielectrics (for details,
see \cite{13a} and references therein).

The linewidths of the resonances are determined by
radiative losses associated with the input-output coupling
and losses due to material absorption. For small linewidths,
the total width can be regarded as being the sum of
a purely radiative term and a purely absorptive term.
In that case we may write
\begin{equation}
\label{e5b}
\delta_{tot} \simeq \delta_{rad} + \delta_{abs},
\end{equation} 
where $\delta_{rad}$ corresponds to the linewidth obtained when the
imaginary part of the permittivity is set equal to zero.
   

\subsection{Resonances below the band gap}
\label{sec2.1}

We first consider the (transverse) field resonances below
the band gap, \mbox{$\omega$ $\!<$ $\!\omega_T$}, where WG waves 
can be observed. These resonances are commonly classified by means   
of three numbers \cite{1,2}: the angular momentum number $l$,
the azimuthal number $m$, and the number $i$ of radial maxima of
the field inside the sphere. In the case of a uniform sphere,
the WG waves are ($2l$ $\!+$ $\!1$)-fold degenerate, i.e.,
the \mbox{$2l$ $\!+$ $\!1$} azimuthal resonances
belong to the same frequency $\Omega_{l,i}$.  

When the imaginary part of the refractive index is much smaller
than the real part, $n_I(\omega)$ $\!\ll$ $\!n_R(\omega)$,
then the method used in Ref.~\cite{14} for solving 
Eqs.~(\ref{e3}) and (\ref{e4}) in the case of
constant, real refractive
index formally applies also to the case of
frequency-dependent, complex refractive   
index. Interpreting the WG waves as resulting from total
internal reflection, it follows that \mbox{$R{\rm Re}\,k_2$
$\gtrsim$ $\!\nu$ $\!\gtrsim$ $\!Rk_1$},
where \mbox{$\nu$ $\!=$ $\!l+1/2$}, and $\nu$ can be assumed to be
large in general. Following Ref.~\cite{14}, the
complex roots $\omega_{l,i}$ of Eqs.~(\ref{e3}) and (\ref{e4})
can then formally given by
\begin{equation}
\label{e5}
\omega_{l,i} = f_{l,i}[n(\omega_{l,i})] + O(\nu^{-1}),
\end{equation}
where
\begin{eqnarray}
\label{e5a}
\lefteqn{
       f_{l,i}[n(\omega_{l,i})] = {c\over Rn(\omega_{l,i})}
       \Biggl\{ \nu + 2^{-1/3}\alpha_i\nu^{1/3}
}
\nonumber\\[.5ex]&&\hspace{2ex}
       -{P\over \left[ n(\omega_{l,i})^2 - 1 \right]^{1/2} }
       + {3\over10}\,2^{-2/3} \alpha_i^2 \nu^{-1/3}
\nonumber\\[.5ex]&&\hspace{2ex}
       -{2^{-1/3}P \left[ n(\omega_{l,i})^2 -2P^2/3 \right]
       \over \left[ n(\omega_{l,i})^2 - 1 \right]^{3/2} }
       \,\alpha_i\nu^{-2/3}
       \Biggr\},
\end{eqnarray}
with $P$ $\!=$ $\!n(\omega_{l,i})$
and $P$ $\!=$ $\!1/n(\omega_{l,i})$, respectively, for TE
and TM waves, and the $\alpha_i$ are the roots of the Airy
function Ai$(-z)$.
Although Eq.~(\ref{e5}) is not yet an explicit 
expression for the roots as in the case of a
nondispersing and nonabsorbing (idealized) medium, it
is numerically more tractable than the original
equations and offers a first insight 
into the WG waves.


\subsubsection{Frequency-independent permittivity}
\label{sec2.1.2}

In many cases in practice, it may be a good approximation to
ignore the dependence on frequency of the refractive index within a 
chosen frequency interval. {F}rom Eqs.~(\ref{e5}) and (\ref{e5a}),
the positions of the WG waves are then found to be
\begin{equation}
\label{e6}
       \Omega = f(n_R) + O 
       \bigg[ \left(n_I\over n_R \right)^2\bigg] \ .
\end{equation}
For notational convenience, here and in the following we drop
the classification indices. Obviously, the first term in
Eq.~(\ref{e6}) provides a very good approximation 
for the positions of the WG waves, provided that the
ratio $n_I/n_R$ is sufficiently small. Note that
\mbox{$n$ $\! \sim$ $\! 1.45$ $\! +$ $\! i 10^{-11}$} for
silica \cite{3} and {$n$ $\! \sim$ $\! 1.47$ $\! +$ $\! i 10^{-7}$}
for glycerol \cite{15} in the  optical region.
In other words, small material absorption does not affect
much the positions of the WG waves.

The contribution to the linewidth of the material absorption
can be evaluated from Eq.~(\ref{e5}) by
first order Taylor expansion of $f(n)$ at $n_R$, 
\begin{equation}
\label{e6.1}
     \delta_{abs} \simeq - n_I f'(n_R) \ .
\end{equation}
To calculate the total line width determined by
both radiative and absorption losses,
one has to go back to the original equations (\ref{e3})
and (\ref{e4}), since the radiative losses are
essentially determined by the (disregarded) term
$O(\nu^{-1})$ in Eq.~(\ref{e5}). First-order Taylor expansion of
$M(\omega)$ around $\Omega$ yields
\begin{equation}
\label{e7}
      \delta_{tot} \simeq {\rm Im}
      \left[ {M(\Omega) \over M'(\Omega)} \right],
\end{equation}
where
\begin{equation}
\label{e8}
      M'(\Omega) \simeq  \epsilon-1
\end{equation}        
for TE waves, and
\begin{equation}
\label{e8a}
      M'(\Omega) \simeq (\epsilon-1)
      \left[\left(1+\frac{1}{\epsilon}\right)
           \left(\frac{c\nu}{R\Omega}\right)^2 -1
      \right]
\end{equation}
for TM waves. The radiative contribution to the line\-width
can then calculated by means of Eq.~(\ref{e5b}).

\begin{figure}[!t!]
\noindent
\begin{center}
\epsfig{figure=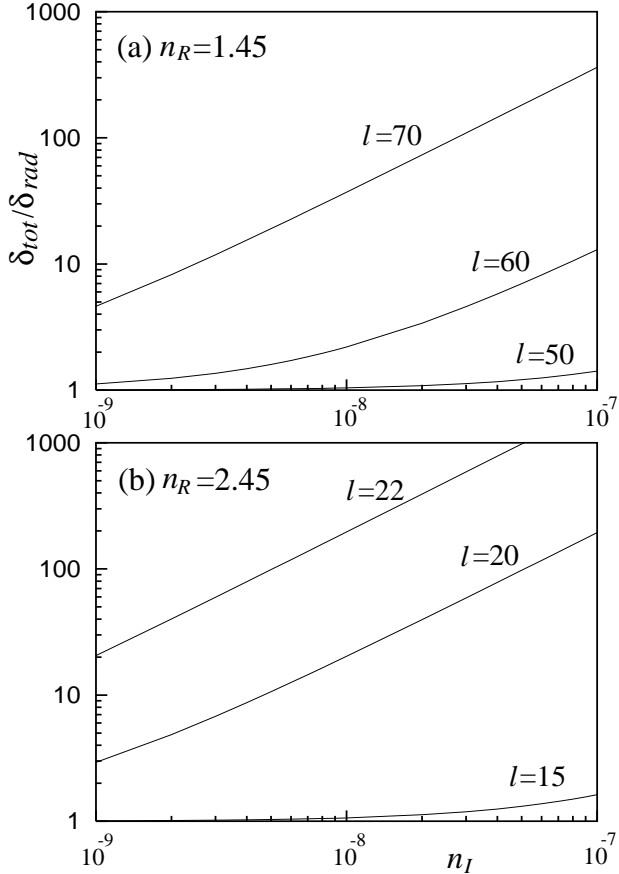,width=1.\linewidth}
\end{center}
\caption{ 
The ratio of the total width $\delta_{tot}$ and the radiative width
$\delta_{rad}$ of TM$_{l,1}$ WG resonances 
is shown as a function of the imaginary part $n_I$
of the refractive index for two values of
the real part $n_R$ of the refractive index
and various values of $l$.
The values
\mbox{$l$ $\!=$ $\!70,\,60,\,50$} correspond to 
\mbox{$R/\lambda$ $\!\simeq$ $\!8.4,\,7.3,\,6.2$}, respectively,
and the values
\mbox{$l$ $\!=$ $\!22,\,20,\,15$} correspond to
\mbox{$R/\lambda$ $\!\simeq$ $\!1.8,\,1.6,\,1.3$}, respectively.
}
\label{f1}
\end{figure}

{F}rom Fig.~\ref{f1} it is seen 
that the contribution to the linewidth of the material absorption
increase with the imaginary part of the refractive index 
$n_I$ and the radius $R$ of the sphere. It is seen that when
the values of $n_I$ and $R$ rise above some threshold values,
then the absorption losses start to dominate the
radiative losses. It is further seen that for chosen $n_I$
the threshold value of $R$ decreases with increasing $n_R$. 
For instance, the absorption losses start to dominate 
the radiative losses if \mbox{$R$ $\!\gtrsim$ $\! 8\lambda$} 
for \mbox{$n$ $\!=1.45$ $\!+$ $\!i10^{-9}$} [see Fig. \ref{f1}(a)], 
and if \mbox{$R$ $\!\gtrsim$ $\! 1.6\lambda$} for \mbox{$n$
$\!=$ $\!2.45+i10^{-9}$} [see Fig. \ref{f1}(b)].
It should be noted, that the approximate results based on  
Eqs.~(\ref{e6.1}) and (\ref{e7}) are in good agreement with
the exact ones (without dispersion). 
%

Introducing the quality factor $Q$ $\!=$ $\Omega/(2\delta)$, 
the ab\-sorp\-tion-assisted part 
$Q_{abs}$ $\!=$ $\!\Omega/(2\delta_{abs})$
is derived from Eqs.~(\ref{e5a})--(\ref{e6.1}) to be
\begin{eqnarray}
\label{e9}
      Q_{abs} = {n_R\over 2n_I}
      + {n_R\over 2n_I}{(2n_R-P) \over 
         (n_R^2-1)^{3/2} } \,\nu^{-1}
      + O\bigl(\nu^{-5/3}\bigr)
\end{eqnarray}
[$P$ $\!=$ $\!n_R$ $(1/n_R)$ for TE (TM) waves].
When the absorption losses dominate the radiative ones,
it is frequently assumed \cite{3,16} that the quality factor
is simply given by the ratio of the (plane-wave) absorption
length and the wavelength, i.e., $Q$ $\!=$
$\!n_R/(2n_I)$, which is just 
the first term on the right-hand side in Eq.~(\ref{e9}).
As long as the values of $\nu$ and/or $n_R$ are large,
which is the case for WG waves, 
the second term on the right-hand side in Eq.~(\ref{e9})
is typically less than a few percent, and the first term is
indeed the leading term. Nevertheless, Eq.~(\ref{e9})
may become significantly wrong, because dispersion
is fully disregarded. Moreover, the radiative losses
can dominate material absorption. 


\subsubsection{Frequency-dependent permittivity}
\label{sec2.1.3}

In order to obtain the exact quality factor in dependence
on the resonance frequency, the permittivity as a function of
frequency must be known and the calculations should be based
on the original equations (\ref{e3}) and (\ref{e4}).
In particular, the quality factor $Q_{rad}$,
which accounts for the line broadening associated with the input--output
coupling, can be obtained by setting the imaginary part of the
permittivity equal to zero,
so that \mbox{$Q$ $\!=$
$\!Q_{rad}$}. Recalling Eq.~(\ref{e5b}), we may write
\mbox{$1/Q$ $\!=$ $1/Q_{tot}$ $\!=$ $\!1/Q_{rad}$ $\!+$ $\!1/Q_{abs}$},
from which the quality factor $Q_{abs}$, which accounts for the 
line broadening due to material absorption, can then be
calculated. Typical results obtained for TM WG waves
on the basis of the model permittivity in Eq.~(\ref{e10})
are shown in Fig.~\ref{f12}. 

{F}rom Fig.~\ref{f12}(a) it is seen that (for chosen radius of
the microsphere) $Q_{abs}$ decreases with increasing frequency 
(i.e., increasing $n_I$ $\!\sim$ $\!\epsilon_I$),
whereas $Q_{rad}$ increases with increasing frequency 
(i.e., increasing $n_R$ $\!\sim$ $\!\sqrt{\epsilon_R}$).
Sufficiently below the band gap where $n_R$ is small
and the radiative losses dominate, $Q_{tot}$ follows $Q_{rad}$.
With increasing frequency the absorption losses
increase and can eventually dominate the radiative losses,  
so that now $Q_{tot}$ follows $Q_{abs}$.

In Fig.~\ref{f12}(b), the values of $Q_{abs}$ calculated
from the exact equation (\ref{e4}) are compared with
the approximate values according to Eq.~(\ref{e9}).
It is seen that the neglect of dispersion in Eq.~(\ref{e9})
leads to some underestimation of the absorption-assisted
quality factor. The difference between the exact and the
approximate values increase with frequency, because of
the increasing dispersion.
We have also calculated $Q_{abs}$ from the approximate
equation (\ref{e5}) [together with Eq.~(\ref{e5a})]
and found a good agreement with the exact result.

Figure \ref{f12}(c) illustrates the influence on the
quality factor of the radius of microsphere and the
strength of absorption. As expected, for lower frequencies
where $Q_{tot}$ follows $Q_{rad}$, the quality is
improved if the radius of the sphere is increased.
For higher frequencies where $Q_{tot}$ follows
$Q_{abs}$, the quality becomes independent of the
radius. Note that with increasing radius the
(frequency) distance between neighboring resonances
decreases.  


\subsection{Resonances inside the band gap}
\label{sec2.2}

Let us now look for (transverse) field resonances inside
the band gap, which is a strictly forbidden zone for nonabsorbing
bulk material.
Inside the gap, we may assume that the real part of the
refractive index is much smaller than the imaginary part,
\mbox{$n_R(\omega)$ $\! \ll$ $\!n_I(\omega)$}. 
Using Bessel-function expansion (Appendix B), 
it can be shown that there are no TE resonances
[Eq.~(\ref{e3}) has no solutions], and
the complex roots of Eq.~(\ref{e4}), which determine the
TM resonances, can formally be given by
%
\begin{figure}[!t!]
\noindent
\begin{center}
\epsfig{figure=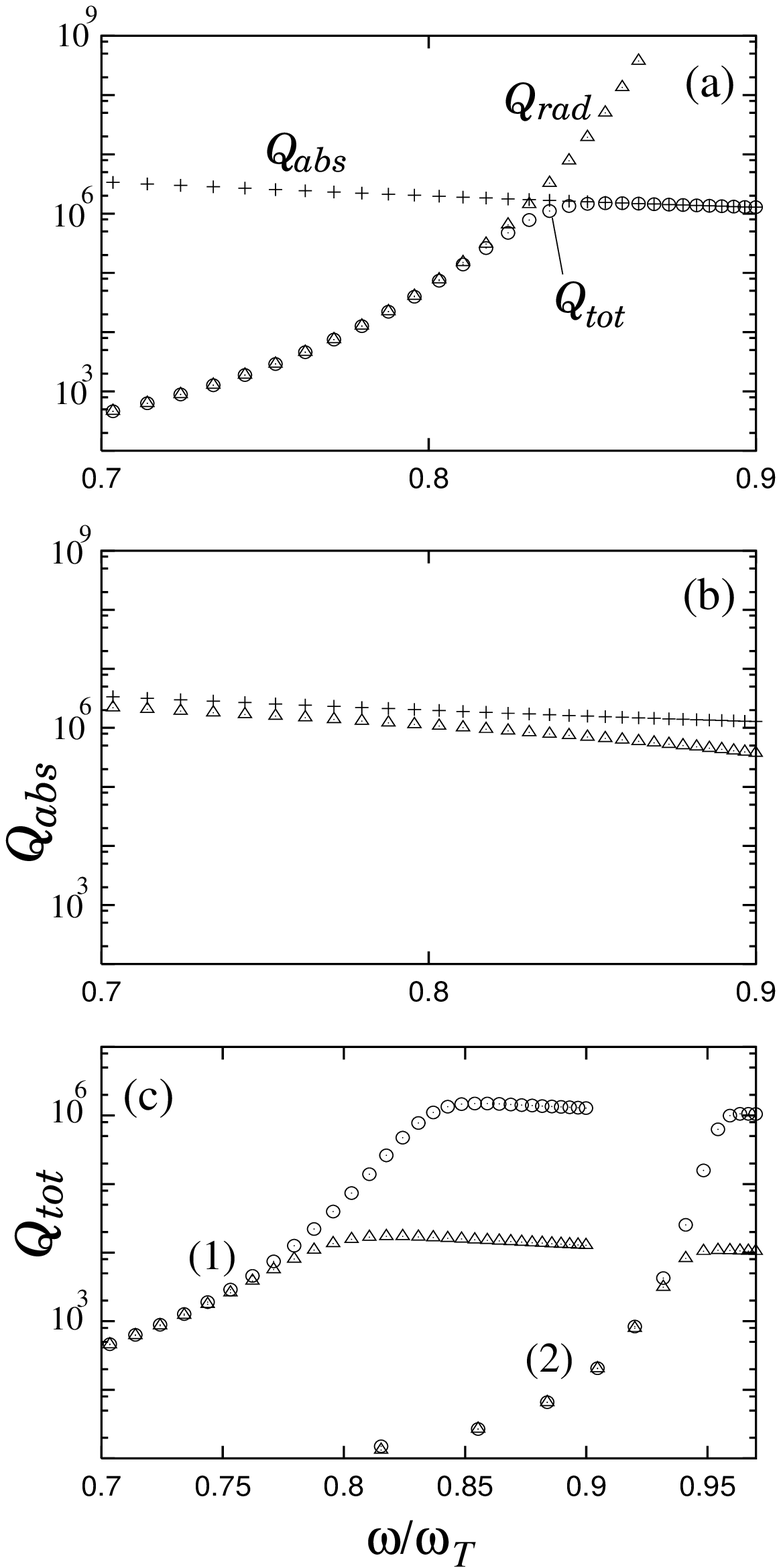,width=1.\linewidth}
\end{center}
\caption{
Quality factors of TM$_{l,1}$ WG field resonances
in a dielectric microsphere of radius $R$ and complex
permittivity $\epsilon(\omega)$ according to
Eq.~(\protect\ref{e10}) \mbox{($\omega_P$ $\!=$ $\!0.5\,\omega_T$)}.
(a) Exact values of $Q_{rad}$, $Q_{abs}$, and $Q_{tot}$ for
\mbox{$R$ $\!=$ $\!10\,\lambda_T$},
\mbox{$\gamma$ $\!=$ $\!10^{-6}\omega_T$},
\mbox{$48$ $\!\leq$ $\!l$ $\!\leq$ $\!78$}.
\mbox{(b) E}xact values of $Q_{abs}$ (+) and approximate
values ($\triangle$) according to Eq.~(\protect\ref{e9})
for the same parameters as in (a).
(c) Exact values of $Q_{tot}$
for \mbox{$\gamma/\omega_T$ $\!=$ $\!10^{-6}$} ({\large$\circ$})
and \mbox{$\gamma/\omega_T$ $\!=$ $\!10^{-4}$} ($\triangle$);
(1) \mbox{$R$ $\!=$ $\!10\,\lambda_T$},
\mbox{$48$ $\!\leq$ $\!l$ $\!\leq$ $\!78$};
\mbox{(2) $R$ $\!=$ $\!2\,\lambda_T$},
\mbox{$10$ $\!\leq$ $\!l$ $\!\leq$ $\!22$}. 
}
\label{f12}
\vspace*{5mm}
\end{figure}
%
\begin{equation}
\label{e11}
     \omega_l = {c\over R} 
     \Biggl[ \nu\sqrt{1\!+\!{1\over \epsilon(\omega_l)}}
     + {\epsilon(\omega_l)^2\!+\!\epsilon(\omega_l)\!+\!1 
        \over 2\epsilon(\omega_l)
        \sqrt{-\epsilon(\omega_l)\!-\!1} }
     \Biggr]
     + O\bigl(\nu^{-1}\bigr),
\end{equation}
where the condition 
%
\begin{figure}[!t!]
\noindent
\begin{center}
\epsfig{figure=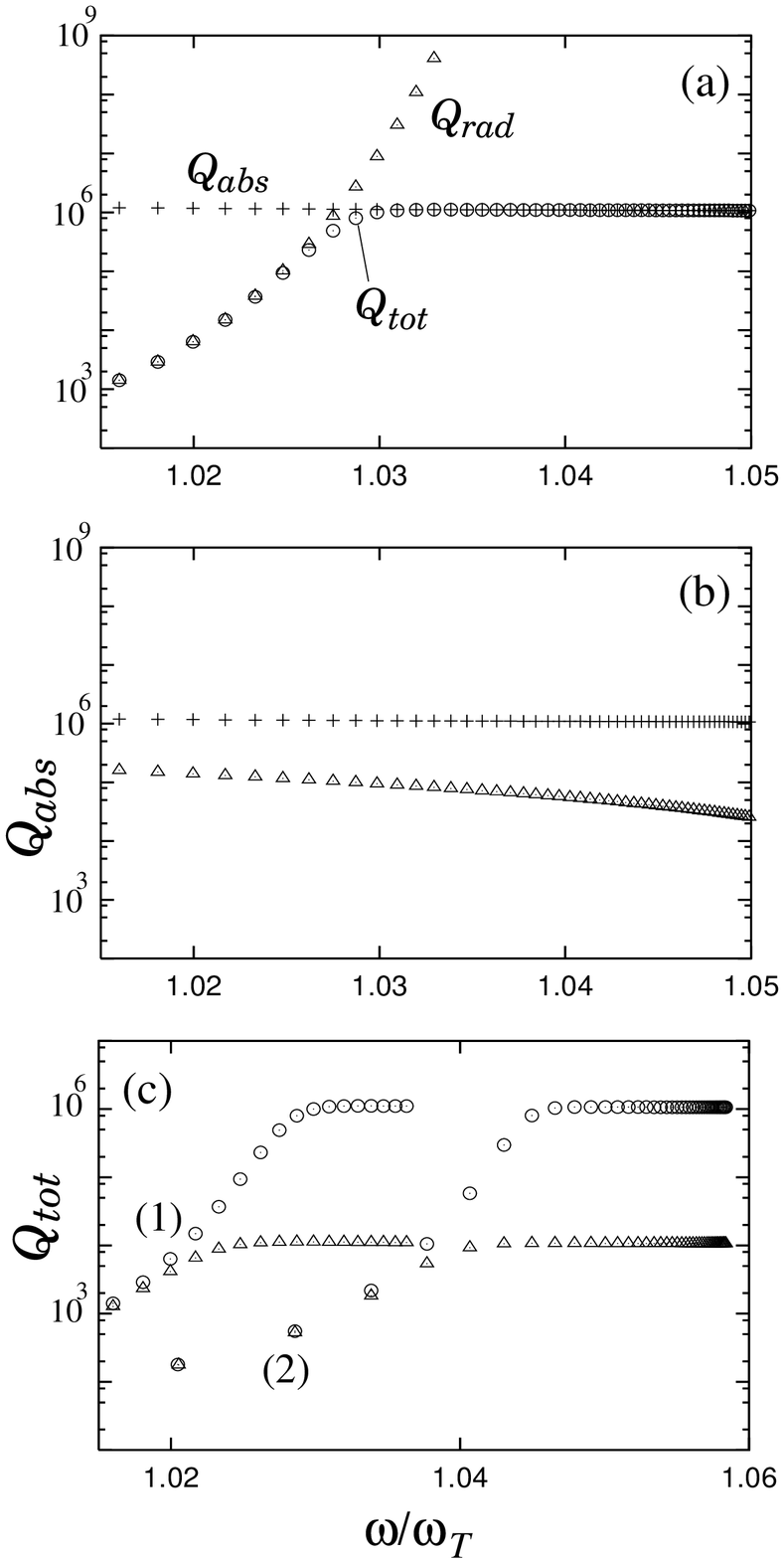,width=1.\linewidth}
\end{center}
\caption{
Quality factors of TM$_l$ SG field resonances in 
a dielectric microsphere of radius $R$ and complex 
permittivity $\epsilon(\omega)$ according to
Eq.~(\protect\ref{e10}) ($\omega_P$ $\!=$ $\!0.5\,\omega_T$).
(a) Exact values of $Q_{rad}$, $Q_{abs}$, and $Q_{tot}$ for
\mbox{$R$ $\!=$ $\!10\,\lambda_T$},
\mbox{$\gamma$ $\!=$ $\!10^{-6}\omega_T$},
\mbox{$70$ $\!\leq$ $\!l$ $\!\leq$ $\!120$}.
\mbox{(b) E}xact values of $Q_{abs}$ (+) and approximate
values ($\triangle$) according to Eq.~(\protect\ref{e12.1})
for the same parameters as in (a).
(c) Exact values of $Q_{tot}$
for \mbox{$\gamma/\omega_T$ $\!=$ $\!10^{-6}$} ({\large$\circ$})
and \mbox{$\gamma/\omega_T$ $\!=$ $\!10^{-4}$} ($\triangle$);
(1) \mbox{$R$ $\!=$ $\!10\lambda_T$},
\mbox{$70$ $\!\leq$ $\!l$ $\!\leq$ $\!86$};
\mbox{(2) $R$ $\!=$ $\!2\,\lambda_T$},
\mbox{$15$ $\!\leq$ $\!l$ $\!\leq$ $\!55$}. 
The gap is from $\omega_T$
to $\omega_L\simeq 1.12\,\omega_T$ [cf. Fig.\protect\ref{f1a}]. 
According to Eq.~(\protect\ref{e12}), resonances exist for 
\mbox{$\Omega_l$ $\!<$ $\!1.0607\,\omega_T$}.
}
\label{f2}
\vspace*{5mm}
\end{figure}
%
\begin{equation}
\label{e11a}
     \epsilon_R(\omega_l) < -1 
\end{equation}
has been required to be satisfied. Note that for the 
Drude-Lorentz model (\ref{e10}), condition (\ref{e11a})
leads to
\begin{equation}
\label{e12}
     \Omega_l={\rm Re}\, \omega_l <
     \sqrt{\omega_T^2+{\textstyle\frac{1}{2}}\omega_P^2} \, .
\end{equation}

The total linewidth can be determined according to 
Eq.~(\ref{e7}), which is still valid here. 
If we ignore the dependence on frequency of the permittivity,
then the expression on the right-hand side of Eq.~(\ref{e11})
can be regarded as a function of the refractive index.  
The contribution to the linewidth of the material absorption
can then be calculated by first-order Taylor expansion at $n_I$
of this function in close analogy to Eq.~(\ref{e6.1}) (with
the roles of $n_R$ and $n_I$ being exchanged).
In terms of the quality factor \mbox{$Q_{abs}$ $\!=$
$\!\Omega/(2\delta_{abs})$} the result reads
\begin{equation}
\label{e12.1}
Q_{abs} = \frac{n_I(n_I^2\!-\!1)}
         {2n_R} + O\bigl(\nu^{-1}\bigr).
\end{equation}
Since now $n_R$ is proportional to $\gamma$, 
$Q_{abs}$ is again proportional to $\gamma^{-1}$.

It is well known that when the condition (\ref{e11a}) is satisfied,
then surface-guided (SG) waves can be excited -- waves
that are bound to the interface and whose amplitudes are 
damped into either of the neighboring media \cite{17a}.
Typical examples are surface phonon polaritons for dielectrics and
surface plasmon polaritons for metals. Obviously, the TM resonances
as determined by Eq.~(\ref{e11}) correspond to SG waves of a sphere.
Note that for the SG waves, in contrast to the WG waves,
each angular momentum number $l$ is associated with only 
one wave.

Examples of
the frequencies and quality factors of SG 
waves are plotted in Fig.~\ref{f2}.
{F}rom a comparison with Fig.~\ref{f12} it is seen that
the quality factors of the SG waves are comparable with
those of the WG waves and also behave in a similar way
as the latter. So, the radiative losses are the dominant ones
for lower-order resonances, whereas for higher-order resonances
the losses essentially arise from material absorption
[Figs.~\ref{f2}(a) and (c)], and the radiative losses
can be reduced by increasing the radius of the sphere
[Fig.~\ref{f2}(c)]. Further, a neglect of
dispersion again leads to some overestimation
of material absorption [Fig.~\ref{f2}(b)].


\subsection{Ground-state fluctuations}
\label{sec2.3}

When an excited atom is in free space, then its
coupling to the vacuum-field fluctuation gives
rise to spontaneous decay.
In the presence of dielectric bodies,
the fluctuation of the electromagnetic field with respect to
the ground-state of the combined
system that consists of the radiation field
and the dielectric matter must be considered. 
The electric-field correlation in the ground state 
can be characterized by the correlation function 
%
\begin{figure}[!t!]
\noindent
\begin{center}
\epsfig{figure=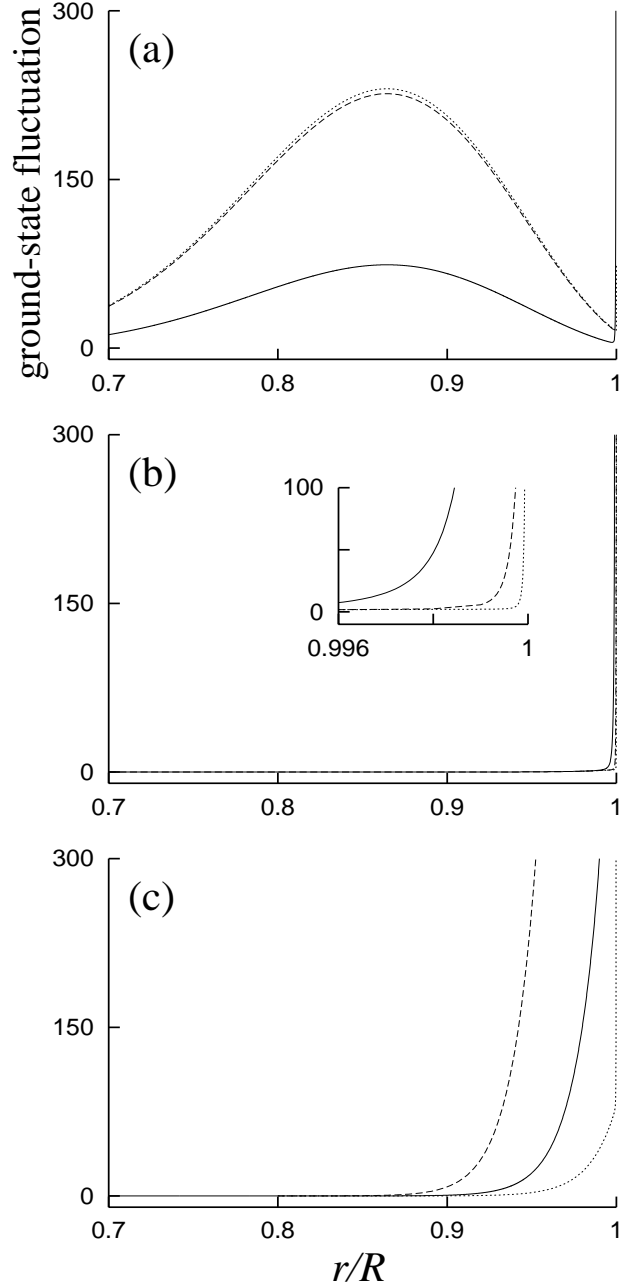,width=1\linewidth}
\end{center}
\caption{ 
Spatial variation of the
ground-state fluctuations
$P_{rr}({\bf r},\omega)$ (in arbitrary units)
inside a  dielectric microsphere of radius
\mbox{$R$ $\!=$ $\!2\,\lambda_T$} and complex 
permittivity $\epsilon(\omega)$ according to
Eq.~(\protect\ref{e10})
[\mbox{$\omega_P$ $\!=$ $\!0.5\,\omega_T$}; 
\mbox{$\gamma/\omega_T$ $\!=$ $\!10^{-4}$} (solid line),
$10^{-6}$ (dashed line), and  $10^{-8}$ (dotted line)].
(a) \mbox{$\omega$ $\!=$ $\!0.94042\,\omega_T$}, 
TM$_{16,1}$ WG field resonance.
(b) \mbox{$\omega$ $\!=$ $\!1.02811\,\omega_T$},  
TM$_{16}$ SG field resonance.   
(c) \mbox{$\omega$ $\!=$ $\!1.05339\,\omega_T$},  
TM$_{30}$ SG field resonance. 
}
\label{f21}
\vspace*{5mm}
\end{figure}
%
\begin{eqnarray}
\label{e12a}
\lefteqn{
      \langle 0| \underline{\hat{E}}_i({\bf r},\omega)
      \underline{\hat{E}}^\dagger_j({\bf r}',\omega')
      |0\rangle   
}
\nonumber\\ &&\hspace{2ex}
      = {\hbar\omega^2\over \pi\epsilon_0 c^2}
      {\rm Im}\ G_{ij}({\bf r},{\bf r}',\omega)
      \delta(\omega-\omega') \ ,
\end{eqnarray}
%
\begin{figure}[!t!]
\noindent
\begin{center}
\epsfig{figure=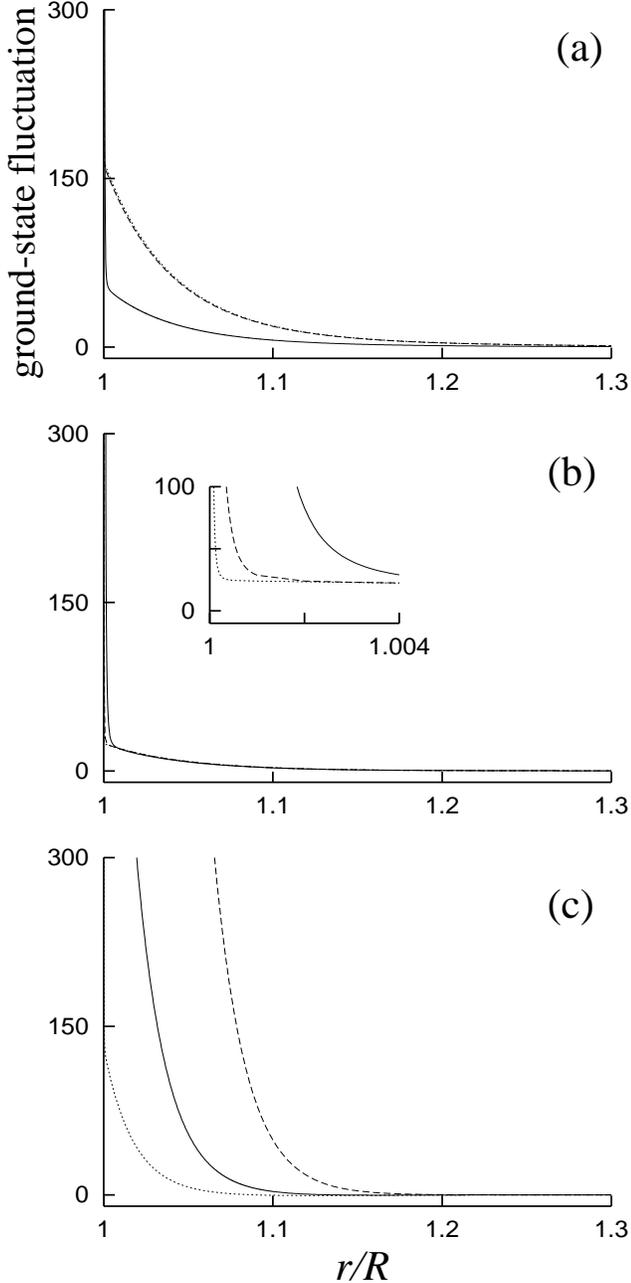,width=1\linewidth}
\end{center}
\caption{ 
The same as in Fig.~\protect\ref{f21}, but for outside 
the sphere.
}
\label{f22}
\end{figure}
%
\noindent
where $\underline{\hat{E}}_i({\bf r},\omega)$ is the
electric-field operator in the frequency domain and 
$G_{ij}({\bf r},{\bf r}',\omega)$ is the classical Green
tensor of the dielectric-matter formation of given   
complex permittivity $\epsilon({\bf r},\omega)$ that
satisfies the Kramers-Kronig relations.
For equal spatial arguments, 
the power spectrum of the ground-state fluctuation can be obtained
according to the relation
\begin{equation}
\label{e12b}
      \langle 0| \underline{\hat{E}}_i({\bf r},\omega)
      \underline{\hat{E}}^\dagger_j({\bf r},\omega')
      |0\rangle   =
      P_{ij}({\bf r},\omega)
      \delta(\omega-\omega'),
\end{equation}
which together with Eq.~(\ref{e12a}) reveals that
\begin{equation}
\label{e12c}
      P_{ij}({\bf r},\omega) =
      {\hbar\omega^2\over \pi\epsilon_0 c^2}\,
      {\rm Im}\ G_{ij}({\bf r},{\bf r},\omega).
\end{equation}

{F}rom the linearity of the Maxwell equations it follows
that the Green tensor for a dielectric body
can be decomposed as [cf. Eq.~(\ref{A1})]
\begin{equation}
\label{e12d}
      G_{ij}({\bf r},{\bf r}',\omega) = 
      G_{ij}^0({\bf r},{\bf r}',\omega) + 
      G_{ij}^R({\bf r},{\bf r}',\omega),
\end{equation}
where $G_{ij}^0$ is the Green tensor for either
the vacuum (outside the body) or the bulk material (inside the
body), and the scattering term $G_{ij}^R$ ensures the correct
boundary conditions (at the surface of discontinuity).
Inside the body ${\rm Im}\,G_{ij}^0$ becomes singular
for \mbox{${\bf r}$ $\!\to$ $\!{\bf r}'$}, and some
regularization (e.g., averaging over a small volume) is
required. Since we are only interested in the spatial variation of
the ground-state fluctuation, we may disregard the 
(space-independent) contribution that arises
from ${\rm Im}\,G_{ij}^0$.

The spatial variation of the radial diagonal component
$P_{rr}$ inside and outside the microsphere under
consideration is illustrated in Figs.~\ref{f21} and \ref{f22},
respectively. Note that $P_{rr}$ only refers to
the TM field noise [cf. Eqs. (\ref{A3}) -- (\ref{A5})].
It is seen that the fluctuation associated
with the WG-type field [Figs.~\ref{f21}(a) and \ref{f22}(a)]
essentially concentrates inside the sphere, whereas the
fluctuation associated with the SG-type field [Figs.~\ref{f21}(b,c)
and \ref{f22}(b,c)] concentrates in the vicinity of the surface
of the sphere. In both cases, a singularity is observed at
the surface. In Figs.~\ref{f21}(a) and \ref{f22}(a),
we have restricted our attention to a single-maximum
WG field, i.e., \mbox{$i$ $\!=$ $\!1$}.
Otherwise \mbox{$i$ ($>$ $\!1$)} maxima would be
observed inside the sphere. 

The strong enhancement of the fluctuation which leads to the
singularity at the surface comes from absorption
and gives rise to nonradiative energy transfer
from the atom to the medium 
(see Sec.~\ref{sec3.2}).
Clearly, for sufficiently small absorption the divergence
becomes meaningless, because the required (small) spatial
resolution would be illicit. 
{F}rom Figs.~\ref{f21}(a) and \ref{f22}(a)
it is seen that material absorption reduces
the fluctuation of the WG field inside and outside the sphere,
and Figs.~\ref{f21}(b,c) and \ref{f22}(b,c) show that
absorption changes the area on either
side of the surface over which the SG field fluctuation  
extends. As expected, the effects are less pronounced for
low-order resonances where the radiative losses
are the dominant ones [compare Figs.~\ref{f21}(b) and (c)
with Figs.~\ref{f22}(b) and (c) respectively].


\section{Spontaneous decay}
\label{sec3}

\subsection{Basic formulae}
\label{sec3.1}

In the weak-coupling regime, an excited atom near a
a dispersing and absorbing body decays exponentially,
where the decay rate is given by 
\begin{equation}
\label{e13}
         A = {2k_A^2\mu_i\mu_j\over \hbar\epsilon_0}\, 
         {\rm Im}\,G_{ij}({\bf r}_A,{\bf r}_A,
         \omega_A), 
\end{equation}
($k_A$ $\!=$ $\!\omega_A/c$; $\bbox{\mu}$, atomic 
transition dipole moment),
%
and the contribution of the body to the shift of the transition
frequency reads 
\begin{equation}
\label{e141}
\delta\omega_A
   = {\mu_i\mu_j\over \pi\hbar\epsilon_0}\,
   {\cal P}\!\int_0^\infty d\omega\, {\omega^2\over c^2} 
   \frac{{\rm Im}\,G_{ij}^{R}({\bf r}_A,{\bf r}_A,\omega)}
   {\omega-\omega_A}\,,
\end{equation}
which can be rewritten, on using the Kramers-Kronig
relations, as
\begin{eqnarray}
\label{e14}
\lefteqn{
\delta\omega_A
   = {k_A^2\mu_i\mu_j\over \hbar\epsilon_0}\,
   \biggl[ 
   {\rm Re}\,G_{ij}^R({\bf r}_A,{\bf r}_A,\omega_A)
}
\nonumber\\&&\hspace{2ex}
   -\, {{\cal P}\over \pi}\!\int_0^\infty d\omega\, 
   {\omega^2\over \omega_A^2} 
   \frac{{\rm Im}\,G_{ij}^R({\bf r}_A,{\bf r}_A,\omega)}
   {\omega+\omega_A}
   \biggr]
\end{eqnarray}
(for details, see \cite{12,19a}).
It is not difficult to see that in Eq.~(\ref{e14})
the second term, which is not sensitive to the atomic
transition frequency, 
is small compared to the first one and can therefore be neglected.
Note that the vacuum Lamb shift (see, e.g., \cite{20}) may be
thought of as being already included in the atomic transition frequency.

The intensity of the spontaneously emitted light registered by a pointlike 
photodetector at position ${\bf r}$ and time $t$ reads \cite{12} 
\begin{eqnarray}
\label{e14a}
\lefteqn{
          I({\bf r},t)  =
          \sum_i \biggl| {k_A^2\mu_j \over \pi\epsilon_0}
          \int_0^t dt' \Big[ C_{u}(t')
}          
\nonumber \\ && \hspace{2ex} \times 
          \int_0^\infty d\omega\, 
          {\rm Im}\, G_{ij} ({\bf r},{\bf r}_A,\omega)
           e^{-i(\omega-\omega_A)(t-t')}
          \Big]\biggr|^2 ,
\end{eqnarray}
where $C_{u}(t)$ is the probability amplitude of finding the
atom in the upper state. Note that Eq.~(\ref{e14a}) is
valid for an arbitrary coupling regime. In particular, 
in the weak coupling regime, where 
the Markov approximation applies,
$C_{u}(t')$ can be taken at \mbox{$t'$ $\!=$ $\!t$} and
put in front of the time integral in Eq.~(\ref{e14a}),
with $C_{u}(t)$ being simply the exponential
\begin{equation}
\label{e14b}
      C_{u}(t)=e^{ \left(-A/2 + i\delta\omega_A \right)t} \ .
\end{equation}
Equation (\ref{e14a}) thus simplifies to 
\begin{equation}
\label{e15}
          I({\bf r},t)  \simeq 
          |{\bf F}({\bf r},{\bf r}_A,\omega_A)|^2 
          e^{-At} ,
\end{equation}
where
\begin{eqnarray}
\label{e16}
\lefteqn{
        F_i({\bf r},{\bf r}_A,\omega_A) = 
        -{ik_A^2\mu_j \over \epsilon_0}
        \biggl[ 
        G_{ij}({\bf r},{\bf r}_A,\omega_A)
}
\nonumber\\&&\hspace{2ex}
        -\, {{\cal P}\over\pi} \int_0^\infty d\omega\, 
        \frac{{\rm Im}\,
        G_{ij}({\bf r},{\bf r}_A,\omega)}
        {\omega+\omega_A}
        \biggr].
\end{eqnarray}
Since the second term on the right-hand side of  
Eq.~(\ref{e16}) is small compared to the first one,
it can be omitted, and the spatial distribution
of the emitted light (emission pattern) can be given by,
on disregarding transit time delay,  
\begin{equation}
\label{e16a}
      |{\bf F}({\bf r},{\bf r}_A,\omega_A)|^2
      \simeq \sum_i 
      \biggl| {k_A^2\mu_j \over \epsilon_0} 
      G_{ij}({\bf r},{\bf r}_A,\omega_A) \biggr|^2.
\end{equation}

Material absorption gives rise to nonradiative decay.
Obviously, the rate (\ref{e13}) as the total decay rate
describes both radiative decay and nonradiative decay.
The fraction of emitted radiation energy can be obtained
by integration of $I({\bf r},t)$ with respect to time
and integration over the surface of a sphere
whose radius is much larger than the extension of the
system,
\begin{equation}
\label{e16b}
W = 2c\epsilon_0\int_0^\infty dt \int_0^{2\pi} d\phi
       \int_0^\pi d\theta\,  \rho^2 \sin\theta \,I({\bf r},t) 
\end{equation}
($\bbox{\rho}$ $\!=$ $\!{\bf r}$ $\!-$ $\!{\bf r}_A$). 
The ratio $W/W_0$, where \mbox{$W_0$ $\!=$ $\!\hbar\omega_A$}
is the emitted energy in free space, then gives us a measure of
the emitted energy, and accordingly, \mbox{$1$ $\!-$ $\!W/W_0$}
measures the energy absorbed by the body.


\subsection{Spontaneous decay rate}
\label{sec3.2}

Using Eqs.~(\ref{A1}) -- (\ref{A3}) 
and (\ref{A3b}) -- (\ref{A5}),
the decay rate (\ref{e13}) for a (with respect to the
microsphere) radially oriented transition dipole moment
can be given by
\begin{eqnarray}
\label{e17}
\lefteqn{
      A^\perp = A_0 \biggl\{ 1 
      +{\textstyle{3\over 2}} \sum_{l=1}^\infty l(l+1)(2l+1)
}
\nonumber\\&&\hspace{2ex} \times\;
      {\rm Re}\,\biggl[ {\cal B}^N_l\!(\omega_A)
      \biggl( {h^{(1)}_l({k_Ar_A}) \over {k_Ar_A}} \biggr)^2 \biggr]
      \biggr\},
\end{eqnarray}
and for a tangential dipole it reads
\begin{eqnarray}
\label{e18}
\lefteqn{
      A^\parallel = A_0 \biggl\{ 1
      + {\textstyle{3\over 4}} \sum_{l=1}^\infty (2l+1)
}
\nonumber\\&&\hspace{2ex}\times\;
      {\rm Re}\,\biggl[ {\cal B}^M_l\!(\omega_A)
      \left( h^{(1)}_l({k_Ar_A}) \right)^2
\nonumber\\&&\hspace{2ex} 
      +\,{\cal B}^N_l\!(\omega_A)
      \biggl( { \bigl[{k_Ar_A} h^{(1)}_l({k_Ar_A})\bigr]^\prime 
      \over {k_Ar_A} } \biggr)^2
      \biggr]
      \biggr\},
\end{eqnarray}
where the prime indicates the derivative with
respect to $k_Ar_A$, and
\begin{equation}
\label{e20}
      A_0 = {k_A^3\mu^2 \over 3\hbar\pi\epsilon_0} 
\end{equation}
is the rate of spontaneous emission in free space. 
Note that a radially oriented transition dipole moment only
couples to TM waves, whereas a tangentially oriented dipole
moment couples to both TM and TE waves.
Equations (\ref{e17}) and (\ref{e18}) generalize the results
obtained for nondispersing and nonabsorbing matter whose
resonance frequencies are far from the atomic transition 
frequencies, i.e., \mbox{$\epsilon$ $\!=$
$\epsilon_R$ $\!>$ $\!1$}, \cite{7,8} to arbitrary
Kramers-Kronig consistent matter, without placing restrictions
to the transition frequency.
%
\begin{figure}[!t!]
\noindent
\begin{center}
\epsfig{figure=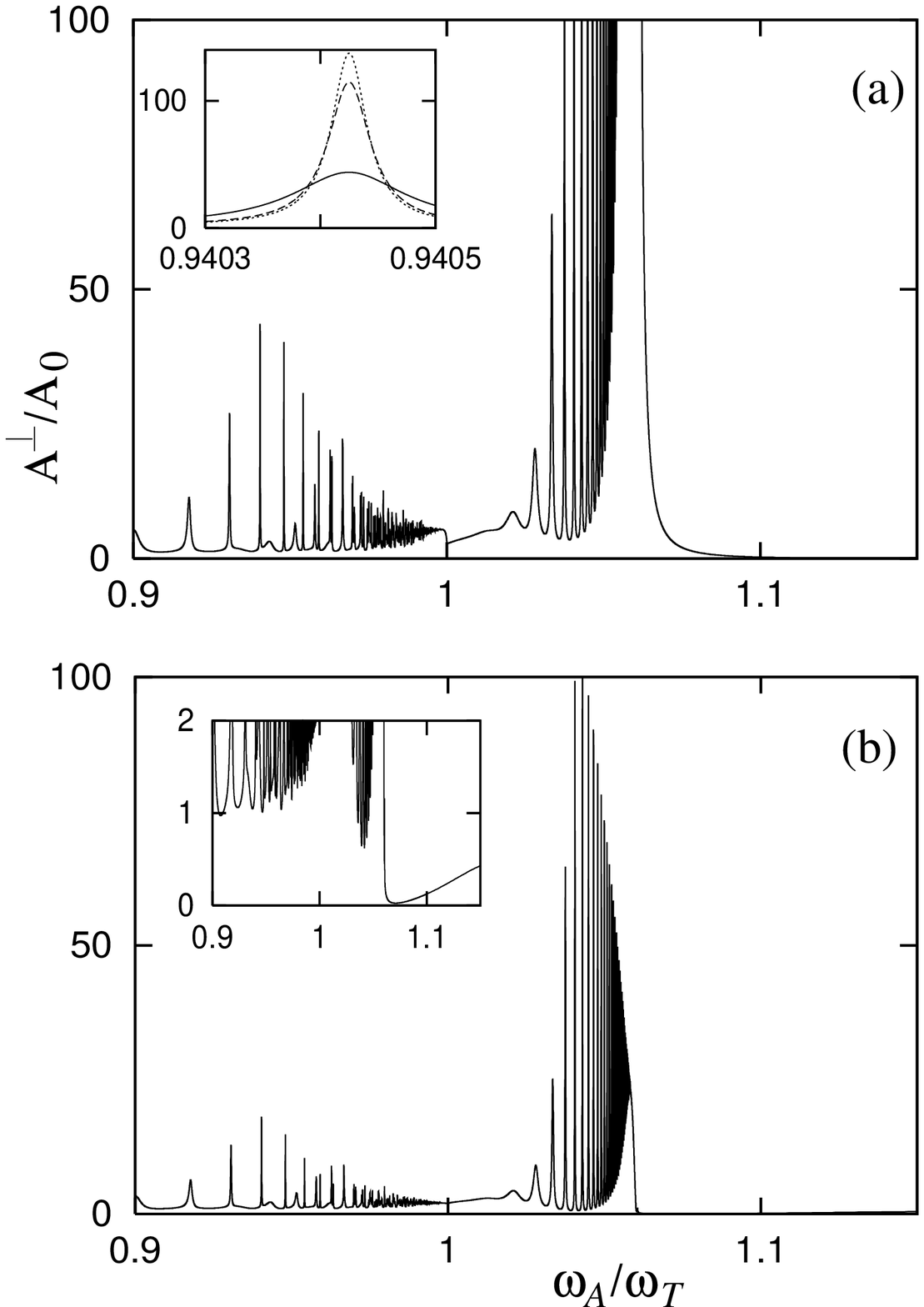,width=1\linewidth}
\end{center}
\caption{ 
Decay rate versus atomic transition frequency 
for a radially oscillating dipole near a
microsphere of radius \mbox{$R$ $\!=$ $\!2\,\lambda_T$}
and complex permittivity $\epsilon(\omega)$ according to
Eq.~(\protect\ref{e10})
(\mbox{$\omega_P$ $\!=$ $\!0.5\,\omega_T$}, 
\mbox{$\gamma/\omega_T$ $\!=$ $\!10^{-4}$}).
(a)
\mbox{$\Delta r$ $\!=$ $\!0.02\,\lambda_T$}; 
inset:  
\mbox{$\gamma/\omega_T$ $\!=$ $\!\!10^{-4}$} (solid line), 
$10^{-5}$ (dashed line), 
$10^{-6}$ (dotted line). 
(b) \mbox{$\Delta r$ $\!=$ $\!0.1\,\lambda_T$}. 
}
\label{f3}
\vspace*{0mm}
\end{figure}
%
When the atom is very close to the microsphere, 
Eqs.~(\ref{e17}) and (\ref{e18}) simplify to 
(see Appendix \ref{appC}) 
\begin{equation}
\label{E21}
     A^\perp = {3A_0\over 4k_A^3} 
     {\epsilon_I(\omega_A)
     \over |\epsilon(\omega_A)+1|^2}
     {1\over (\Delta r)^3} + O(1)
\end{equation}
and
\begin{equation}
\label{E22}
     A^\parallel = {3A_0\over 8k_A^3} 
     {\epsilon_I(\omega_A)
     \over |\epsilon(\omega_A)+1|^2}
     \left[ {1\over (\Delta r)^3}
            + {1\over 2R^2\Delta r} \right]
      + O(1),
\end{equation}
with \mbox{$\Delta r$ $\! =$ $\! r_A$ $\!-$ $\!R$} being the
distance between the atom and the surface of the microsphere.
The terms \mbox{$\sim$ $\!(\Delta r)^{-3}$} and
\mbox{$\sim$ $\!(\Delta r)^{-1}$}
result from 
the TM near-field coupling. Since they are proportional to
$\epsilon_I$, they describe nonradiative decay, i.e.,
energy transfer from the atom to the medium,
and reflect the strong enhancement of the electromagnetic-field
fluctuation as \mbox{$\Delta r$ $\!\to$ $\!0$}
(see Sec.~\ref{sec2.3}).
In particular, the terms \mbox{$\sim$ $\!(\Delta r)^{-3}$} in
Eqs.~(\ref{E21}) and (\ref{E22}) are exactly the same as
in the case when the atom is close to a planar
interface \cite{23,24}. Obviously, nonradiative decay
does not respond sensitively to the actual
radiation-field structure. Since the distance of the
atom from the surface of the microsphere must not
be smaller than interatomic distances (otherwise the
concept of macroscopic electrodynamics fails), the
divergence at the surface (\mbox{$\Delta r$ $\!=$ $\!0$}) 
is not observed.

The dependence of the decay rate (\ref{e17}) on the 
rad\-iat\-ion-field structure, which can be observed
for not too small (large) values of $\Delta r$ ($\epsilon_I$),   
is illustrated in Fig.~\ref{f3} for a radially
oriented transition dipole moment. Since the decay rate
is proportional to the spectral density of final states,
its dependence on the transition frequency mimics the
excitation spectrum of the sphere-assisted radiation field.
As a result, both the WG and SG field resonances can strongly
enhance the spontaneous decay. The enhancement decreases
with increasing distance between the atom and the sphere
[compare Figs.~\ref{f3}(a) an(b)]. 
When material absorption increases, then the resonance lines
are broadened at the expense of the heights and the enhancement
is accordingly reduced [see the inset in Fig.~\ref{f3}(a)].

Figure \ref{f3} reveals that
the SG field resonances, which are the more intense ones
in general, can give rise to a much stronger enhancement of
the spontaneous decay than the WG field resonances.
In particular, with increasing angular-momentum number
the lines of the SG field resonances strongly overlap
and huge enhancement factors can be observed for transition
frequencies inside the band gap [e.g., factors  of the order
of magnitude of $10^4$ for the parameters chosen in
Fig.~\ref{f3}(a)].
When the distance between the atom and the sphere increases, then
the atom rapidly decouples from that part of the field. Thus,  
the huge enhancement of spontaneous decay rapidly reduces and
the interval in which inhibition of spontaneous decay is
typically observed, extends accordingly [see Fig.~\ref{f3}(b)]. 


\subsection{Lamb shift}
\label{sec3.3}

Substituting expressions (\ref{A3}) 
and (\ref{A3b}) -- (\ref{A5}) into
Eq.~(\ref{e14}), we find that the shift of the atomic transition
frequency, which is caused by the presence of the microsphere, is
\begin{eqnarray}
\label{e23}
\lefteqn{
      \delta\omega_A^\perp = -  
      {3A_0\over 4} \sum_{l=1}^\infty l(l+1)(2l+1)
}
\nonumber\\&&\hspace{2ex} \times\; 
      {\rm Im} \biggl\{ {\cal B}^N_l\!(\omega_A)
      \biggl[ {h^{(1)}_l({k_Ar_A}) \over {k_Ar_A}} \biggr]^2
               \biggr\} 
\end{eqnarray}
for a radially oriented transition dipole moment, and 
\begin{eqnarray}
\label{e24}
\lefteqn{
      \delta\omega_A^\parallel = -  
      {3A_0\over 8} \sum_{l=1}^\infty (2l+1)
}
\nonumber\\&&\hspace{2ex}\times\;      
      {\rm Im} \biggl\{ {\cal B}^M_l\!(\omega_A)
      \left[ h^{(1)}_l({k_Ar_A}) \right]^2
\nonumber\\&&\hspace{2ex} 
      +\,{\cal B}^N_l\!(\omega_A)
      \biggl[ {\bigl[{k_Ar_A} h^{(1)}_l({k_Ar_A})\bigr]^\prime 
      \over {k_Ar_A}} \biggr]^2
               \biggr\}
\end{eqnarray}
for a tangentially oscillating dipole.
In particular, when the distance between the atom and the sphere
is very small, then Eqs.~(\ref{e23}) and (\ref{e24}) simplify to
[in close analogy to Eqs.~(\ref{E21}) and (\ref{E22})]
\begin{equation}
\label{e25}
     \delta\omega_A^\perp = -{3A_0\over 16k_A^3} 
     {|\epsilon(\omega_A)|^2-1
     \over |\epsilon(\omega_A)+1|^2}
     {1\over (\Delta r)^3} + O(1) 
\end{equation}
and
\begin{equation}
\label{e26}
     \delta\omega_A^\parallel = -{3A_0\over 32k_A^3} 
     {|\epsilon(\omega_A)|^2\!-\!1
     \over |\epsilon(\omega_A)\!+\!1|^2}
     \left[ {1\over (\Delta r)^3}
            +{1\over 2R^2\Delta r} \right]
      + O(1).
\end{equation}
The terms \mbox{$\sim$ $\!(\Delta r)^{-3}$} and
\mbox{$\sim$ $\!(\Delta r)^{-1}$}
again result from the TM near-field coupling,
and the leading terms agree with those obtained
for a planar interface \cite{24}.  
Note that in contrast to the decay rate, the Lamb shift
diverges for \mbox{$\Delta r$ $\!\to$ $\!0$}
even when \mbox{$\epsilon_I$ $\!=$ $\!0$}.

\begin{figure}[!t!]
\noindent
\begin{center}
\epsfig{figure=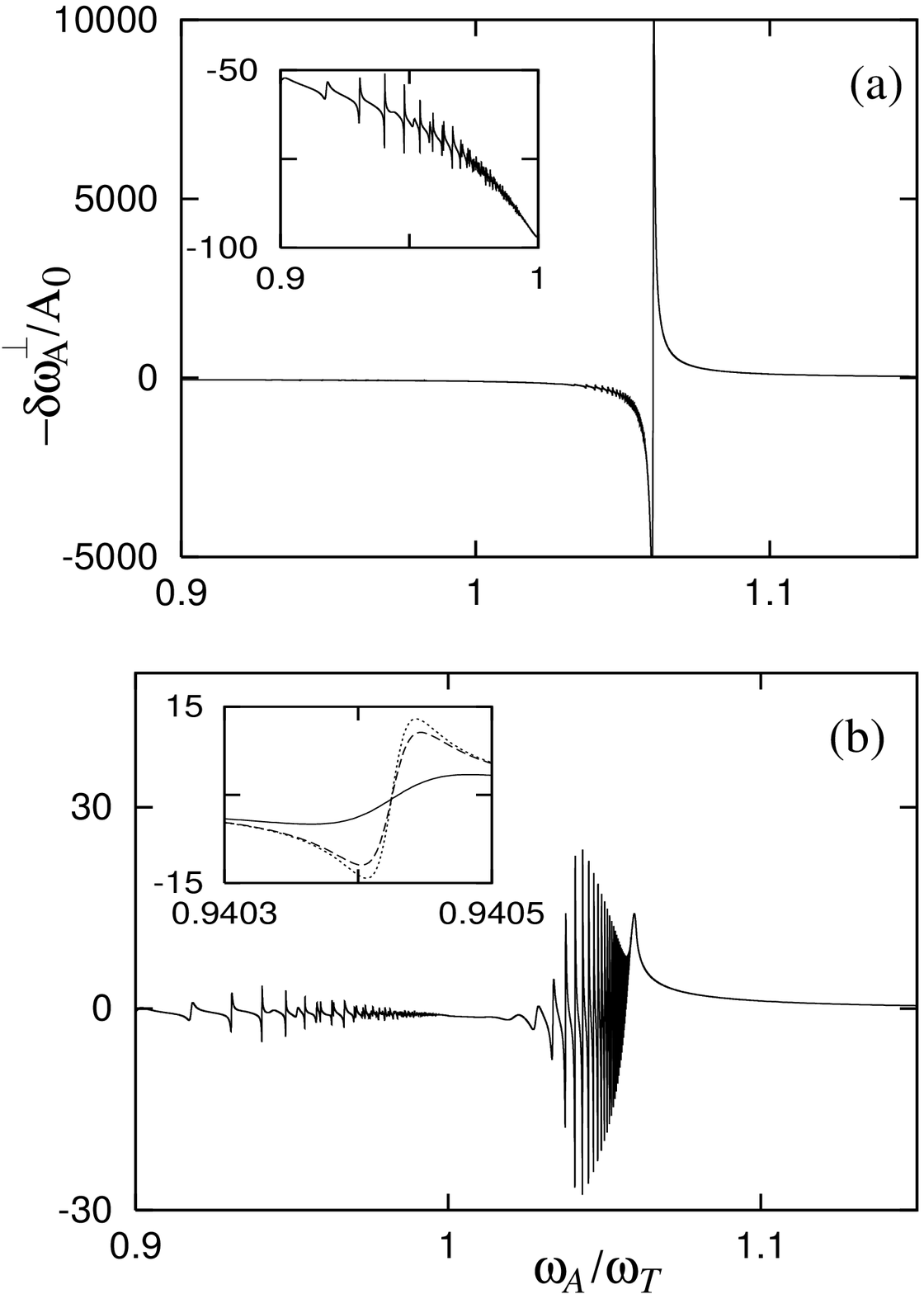,width=1\linewidth}
\end{center}
\caption{
Lamb shift versus atomic transition frequency 
for a radially oscillating dipole near a
microsphere of radius \mbox{$R$ $\!=$ $\!2\,\lambda_T$}
and complex permittivity $\epsilon(\omega)$ according to
Eq.~(\protect\ref{e10})
(\mbox{$\omega_P$ $\!=$ $\!0.5\,\omega_T$}, 
\mbox{$\gamma/\omega_T$ $\!=$ $\!10^{-4}$}).
(a) \mbox{$\Delta r$ $\!=$ $\!0.02\,\lambda_T$}.
\mbox{(b) $\Delta r$ $\!=$ $\!0.1\,\lambda_T$};
inset: \mbox{$\gamma/\omega_T$ $\!=$ $\!\!10^{-4}$} (solid line), 
$10^{-5}$ (dashed line), 
$10^{-6}$ (dotted line). 
}
\label{f5}
\end{figure}

The dependence of the frequency shift (\ref{e23}) on the 
rad\-iat\-ion-field structure for not too small values
of $\Delta r$ is illustrated in Fig.~\ref{f5}.
It is seen 
that each field resonance can give rise to a noticeable
frequency shift in the very vicinity of the
resonance frequency; transition frequencies that are lower (higher) 
than the resonance frequency are shifted to lower (higher)
frequencies. In close analogy to the behavior of the decay rate,
the shift is more pronounced for SG field resonances than
for WG field resonances and can be huge for large angular
momentum numbers when the lines of the SG field resonances
strongly overlap.  

The behavior of the Lamb shift as shown in Fig.~\ref{f5}(b)
can already be seen in the single-resonance limit \cite{24a}. 
If the atomic transition frequency $\omega_A$ is close to a
resonance frequency $\Omega$
of the microsphere, contribution from other 
resonances may be ignored in a first approximation.
Regarding the resonance line as being
a Lorentzian and using contour integration, we obtain from 
Eq.~(\ref{e141})
\begin{eqnarray}
\label{e27}
    \delta\omega_A
    &\simeq& -{A(\Omega) \delta^2           
    \over 4\pi}\, 
    {\cal P}\!\! \int_{-\infty}^\infty \!\!d\omega\, 
    {1\over \omega\!-\!\omega_A}
    {1\over (\omega\!-\!\Omega)^2\!+\!\delta^2}  
\nonumber\\&=&
    -{A(\Omega)\delta \over 2}              
    {\Delta \over \Delta^2+\delta^2} \,,    
\end{eqnarray}    
where 
$A(\Omega)$ is the decay rate as given by Eq.~(\ref{e13})
(with $\Omega$ in place of $\omega_A$), and
\mbox{$\Delta$ $\!=$ $\!\omega_A$ $\!-$ $\!\Omega$}.
In particular,
Eq.~(\ref{e27}) indicates that the frequency shift peaks at
half maximum  on both sides of the resonance line.
With increasing material
absorption, the linewidth
$\delta$
increases while 
$A(\Omega)$
decreases
and thus the absolute values of the frequency shift are
reduced, the distance between the maximum and the minimum
being somewhat increased. 
With decreasing distance between the atom and the microsphere
near-field effects become important and Eq.~(\ref{e27}) fails,
as it can be seen from a comparison of Figs.~\ref{f5}(a) and (b).
%


\subsection{Emitted-light intensity}
\label{sec3.4}

\subsubsection{Spatial distribution}
\label{sec3.4.1}

\begin{figure}[!t!]
\noindent
\begin{center}
\epsfig{figure=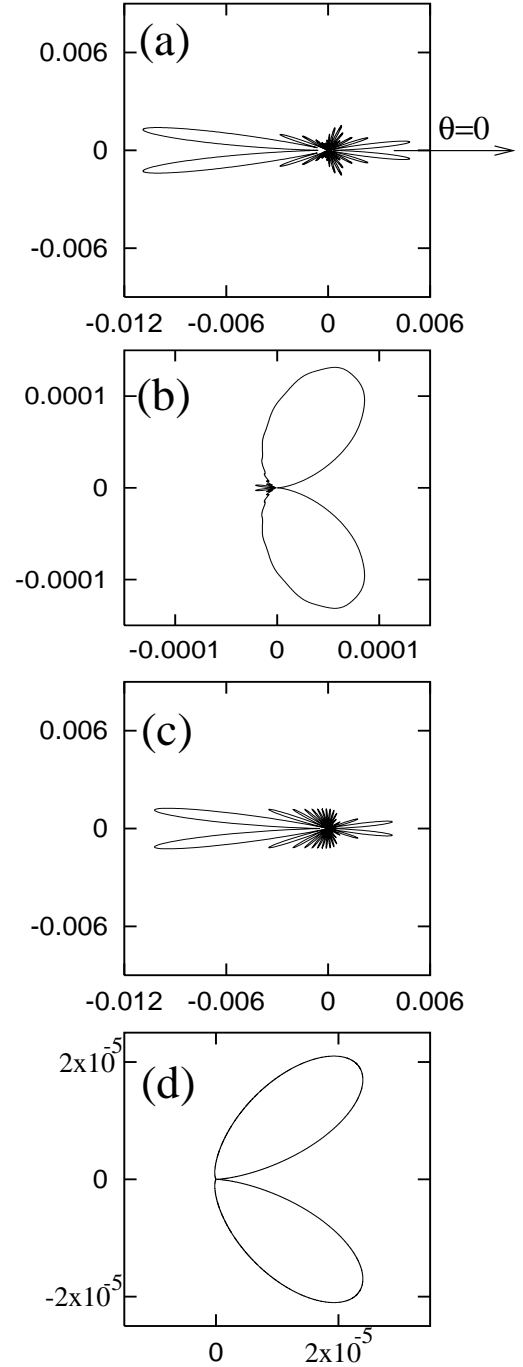,width=.9\linewidth}
\end{center}
\caption{
Polar diagrams of the normalized far-field emission pattern 
$|{\bf F}^\perp({\bf r},{\bf r}_A,\omega_A)|^2/
(k_A^3\mu / 4\pi \epsilon_0)^2$
of a radially oscillating dipole
near a microsphere of radius \mbox{$R$ $\!=$ $\!2\,\lambda_T$}
and complex permittivity $\epsilon(\omega)$ according to
Eq.~(\protect\ref{e10})
(\mbox{$\omega_P$ $\!=$ $\!0.5\,\omega_T$}, 
\mbox{$\gamma/\omega_T$ $\!=$ $\!10^{-4}$},
\mbox{$\Delta r$ $\!=$ $\!0.02\,\lambda_T$},
\mbox{$r$ $\!=$ $\!20\lambda_T$}).
$\omega_A/\omega_T\!=\!$ 
(a) 0.94042, 
(b) 0.999, 
(c) 1.02811, and
(d) 1.06.
}
\label{f6}
\end{figure}

\begin{figure}[!t!]
\noindent
\begin{center}
\epsfig{figure=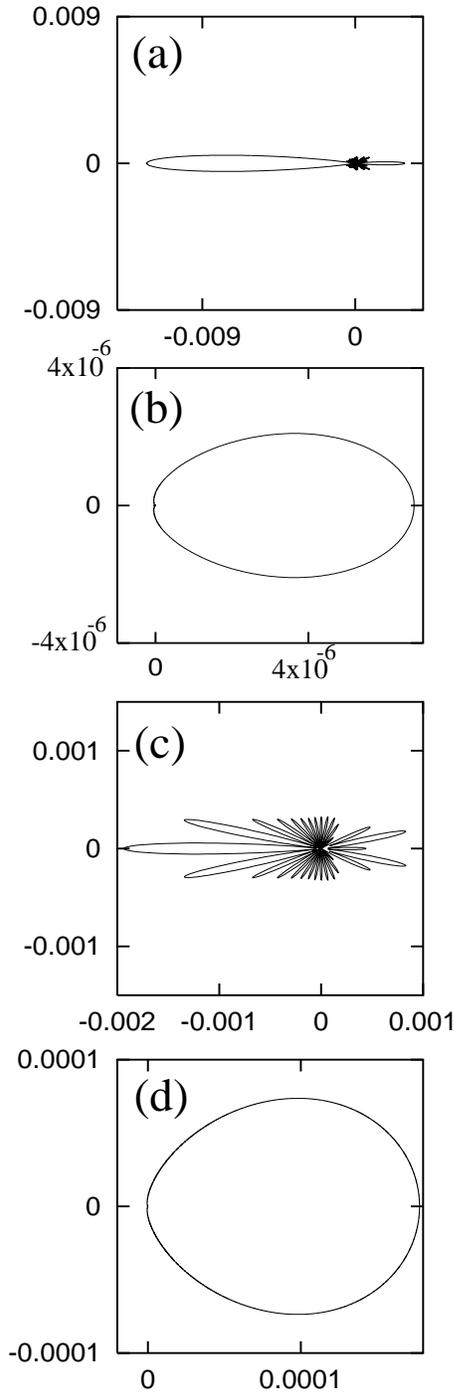,width=.9\linewidth}
\end{center}
\caption{
The same as in Fig.~\protect\ref{f6}, but for a tangentially
oscillating dipole.
}
\label{f7}
\end{figure}

Using Eqs.~(\ref{A1}) -- (\ref{A5}), the function
${\bf F}({\bf r},{\bf r}_A,\omega_A)$, Eq.~(\ref{e16}),
which determines, according to Eq.~(\ref{e15}), the spatial
distribution of the emitted light, can be given by
(\mbox{$\theta_A$ $\!=$ $\!\phi_A$ $\!=$ $\!0$},
\mbox{$r_A$ $\!\le$ $\!r$})
\begin{eqnarray}
\label{e28}
\lefteqn{
     {\bf F}^\perp({\bf r},{\bf r}_A,\omega_A) =
     {k_A^3\mu \over 4\pi \epsilon_0}
     \sum_{l=1}^\infty (2l+1)
}
\nonumber\\&&\hspace{2ex} \times\; 
     {1\over {k_Ar_A}}
     \left[j_l({k_Ar_A}) + {\cal B}_l^N\!(\omega_A)
            h_l^{(1)}({k_Ar_A}) \right]
\nonumber\\&&\hspace{2ex} \times\,
     \biggl[ {\bf e}_r \,l(l\!+\!1)\, {h_l^{(1)}({k_Ar})\over {k_Ar}}
     \,P_l(\cos\theta)
\nonumber\\&&\hspace{4ex} 
     -\, {\bf e}_\theta \,{[{k_Ar}\,h_l^{(1)}({k_Ar})]'\over {k_Ar}}\,
     \sin\theta P_l'(\cos\theta)
     \biggr] 
\end{eqnarray}
for a radially oriented transition dipole moment, and
\begin{eqnarray}
\label{e29}
\lefteqn{
     {\bf F}^\parallel({\bf r},{\bf r}_A,\omega_A) =
     {k_A^3\mu \over 4\pi \epsilon_0}
     \sum_{l=1}^\infty {(2l+1)\over l(l+1)}
}
\nonumber\\&&\times 
     \biggl\{ 
     {\bf e}_r  \cos \phi  
     \tilde{\cal B}_l^N l(l\!+\!1)\,{h_l^{(1)}({k_Ar})\over {k_Ar}}\,
     \sin\theta P_l'(\cos\theta)
\nonumber\\&&\hspace{2ex} 
     + \,{\bf e}_\theta \cos \phi  
     \biggl[\tilde{\cal B}_l^M  h_l^{(1)}({k_Ar}) P_l'(\cos\theta)
\nonumber\\&&\hspace{8ex}   
     + \,\tilde{\cal B}_l^N \,{[{k_Ar}\,h_l^{(1)}({k_Ar})]'\over {k_Ar}}\, 
     \tilde{P}_l(\cos\theta) 
     \biggr] 
\nonumber\\&&\hspace{2ex} 
      -\,{\bf e}_\phi \sin \phi
     \biggl[\tilde{\cal B}_l^M  h_l^{(1)}({k_Ar}) \tilde{P}_l(\cos\theta)
\nonumber\\&&\hspace{8ex}  
     +\,\tilde{\cal B}_l^N \,{[{k_Ar}\,h_l^{(1)}({k_Ar})]'\over {k_Ar}}\, P_l'(\cos\theta)
     \biggr] \biggr\}
\end{eqnarray}
for a tangentially oriented dipole in the $xz$-plane.
Here,
the abbreviating notations
\begin{eqnarray}
\label{e30}
\lefteqn{
     \tilde{\cal B}_l^N = 
     {1\over {k_Ar_A}}
     \Big\{ [{k_Ar_A}j_l({k_Ar_A})]'
}
\nonumber\\&&\hspace{2ex}
     +\, {\cal B}_l^N\!(\omega_A) 
            [{k_Ar_A}h_l^{(1)}({k_Ar_A})]'
     \Big\} ,
\end{eqnarray}
\begin{equation}     
\label{e31}
     \tilde{\cal B}_l^M = 
     j_l({k_Ar_A}) + {\cal B}_l^M\!(\omega_A) h_l^{(1)}({k_Ar_A}) ,
\end{equation}
\begin{equation} 
\label{e32}
     \tilde{P}_l (\cos\theta)=
     l(l+1)P_l(\cos\theta) - \cos\theta P'_l(\cos\theta)
\end{equation}
have been introduced.
Examples of the far-field emission pattern of a radially
oriented transition dipole moment,
$|{\bf F}^\perp({\bf r},{\bf r}_A,\omega_A)|^2$,
and a tangentially oriented transition dipole moment,
$|{\bf F}^\|({\bf r},{\bf r}_A,\omega_A)|^2$,
are plotted in Figs.~\ref{f6} and \ref{f7} respectively.

Let us first restrict our attention to a radially oriented
transition dipole moment. In this case, the far field is
essentially determined by $F^\perp_\theta$, as an inspection
of Eq.~(\ref{e28}) reveals. When the atomic transition frequency
coincides with the frequency of a WG wave of angular momentum
number $l$ far from the band gap [Fig.~\ref{f6}(a)], then
the corresponding $l$-term
in the series (\ref{e28}) obviously
yields the leading contribution to the emitted radiation,
whose angular distribution is significantly determined
by the term \mbox{$\sim$ $\!\sin\theta\, P'_l(\cos\theta)$}.
Since $P_l(\cos\theta)$ is a polynomial with $l$ real, single 
roots in the interval \mbox{$0$ $\!<$ $\!\theta$ $\!<$ $\!\pi$}
\cite{25}, $\sin\theta\,P'_l(\cos\theta)$ must have $l$ extrema
within this interval. Thus, the emission pattern
has $l$ lobes in, say, the $yz$-plane, i.e., $l$
cone-shaped peaks around the $z$-axis, because of symmetry reasons.
The lobes near \mbox{$\theta$ $\!=$ $\!0$} and
\mbox{$\theta$ $\!=$ $\!\pi$} are the most 
dominant ones in general, because of \cite{18}
\begin{equation}
\label{e33}
     -\sin\theta P'_l(\cos\theta) 
     \sim (\sin\theta)^{-1/2} + O\bigr(l^{-1}\bigl)
\end{equation}
($0$ $\!<$ $\!\theta$ $\!<$ $\!\pi$).
Note that the superposition of the leading term with the
remaining terms in the series (\ref{e28}) gives rise
to some asymmetry with respect to the plane
\mbox{$\theta$ $\!=$ $\!\pi/2$}.
%
%
%
When the atomic transition frequency approaches the
band gap (but is still outside it),
a strikingly different behavior is observed [Fig.~\ref{f6}(b)].
The emission pattern changes to a two-lobe 
structure similar to that observed in free space, but bent 
away from the microsphere surface, the emission 
intensity being very small.
Since near the band gap absorption losses dominate,
a photon that is resonantly emitted
is almost certainly absorbed and does not contribute
to the far field in general. If the photon is
emitted in a lower-order WG wave where radiative
losses dominate, it has a bigger chance to escape.
The superposition of
all these weak (off-resonant) contributions
just form the two-lobe emission pattern observed,
as it can also be seen by careful inspection of
the series (\ref{e28}).  

When the atomic transition frequency is inside the band gap and
coincides with the frequency of a SG wave of low order such
that the radiative losses dominate, then the
emission pattern resembles that observed for resonant
interaction with a low-order WG wave [compare Figs.~\ref{f6}(a)
and (c)]. With increasing transition frequency the
absorption losses become substantial and eventually change the
emission pattern in a quite similar way as do below the band gap
[compare Figs.~\ref{f6}(b) and (d)]. Obviously,
the respective explanations are similar in the two cases.      

For a tangentially oriented transition dipole moment
the situation is not lucid in general. Let us
therefore restrict our attention to the far field in
the $xz$-plane, i.e., \mbox{$\phi$ $\!=$ $\!0$} in
Eq.~(\ref{e29}). In this case, the main contribution
to the far field comes from $F^\parallel_\theta$.
The interpretation of the plots in Fig.~\ref{f7}
is quite similar to that of the plots in Fig.~\ref{f6}.
In particular, when the  atomic transition frequency
coincides with a WG field resonance that mainly suffers
from radiative losses, then the $l$ term in the series
(\ref{e29}) that corresponds to the order of the
WG wave is the leading one, and the main contribution to
it stems from the term \mbox{$\sim$ $\!l(l$ $\!+$
$\!1)P_l(\cos\theta)$}. It gives rise to \mbox{$l$ $\!-$ $\!1$}   
lobes in the interval \mbox{$0$ $\!<$ $\!\theta$ $\!<$ $\!\pi$},
and two lobes at 
\mbox{$\theta$ $\!=$ $\!0$} and \mbox{$\theta$ $\!=$ $\!\pi$}, 
which are the most pronounced ones 
[Fig.~\ref{f7}(a)]. These maxima are approximately located at
the positions of the minima of the far field of 
the radially oscillating dipole. Hence, if the 
transition dipole moment has a radial and a tangential
component, a smoothed superimposed field is observed.


\subsubsection{Radiative versus nonradiative decay}
\label{sec3.4.2}

Since both the imaginary part of vacuum Green tensor $G_{ij}^{0}$
and the scattering term $G_{ij}^{R}$ are transverse, 
the decay rate (\ref{e13}) results from the coupling of the
atom to the transverse part of the electromagnetic field.
Nevertheless, the decay of the excited atomic state must not
necessarily be accompanied by the emission of a real photon,
but instead a matter quantum can be created, 
because of material absorption.
To compare the two decay channels,
we have calculated, according to Eq.~(\ref{e16b}),
the fraction $W/W_0$ of the atomic (transition) energy that
is irradiated. Using Eqs.~(\ref{e15}), (\ref{e16b}), and
(\ref{e28}), we derive, on recalling the relations
(\ref{A10}) and (\ref{A11}), for a radially oriented
transition dipole moment 
\end{multicols}
\begin{figure}[!t!]
\noindent
\begin{center}
\epsfig{figure=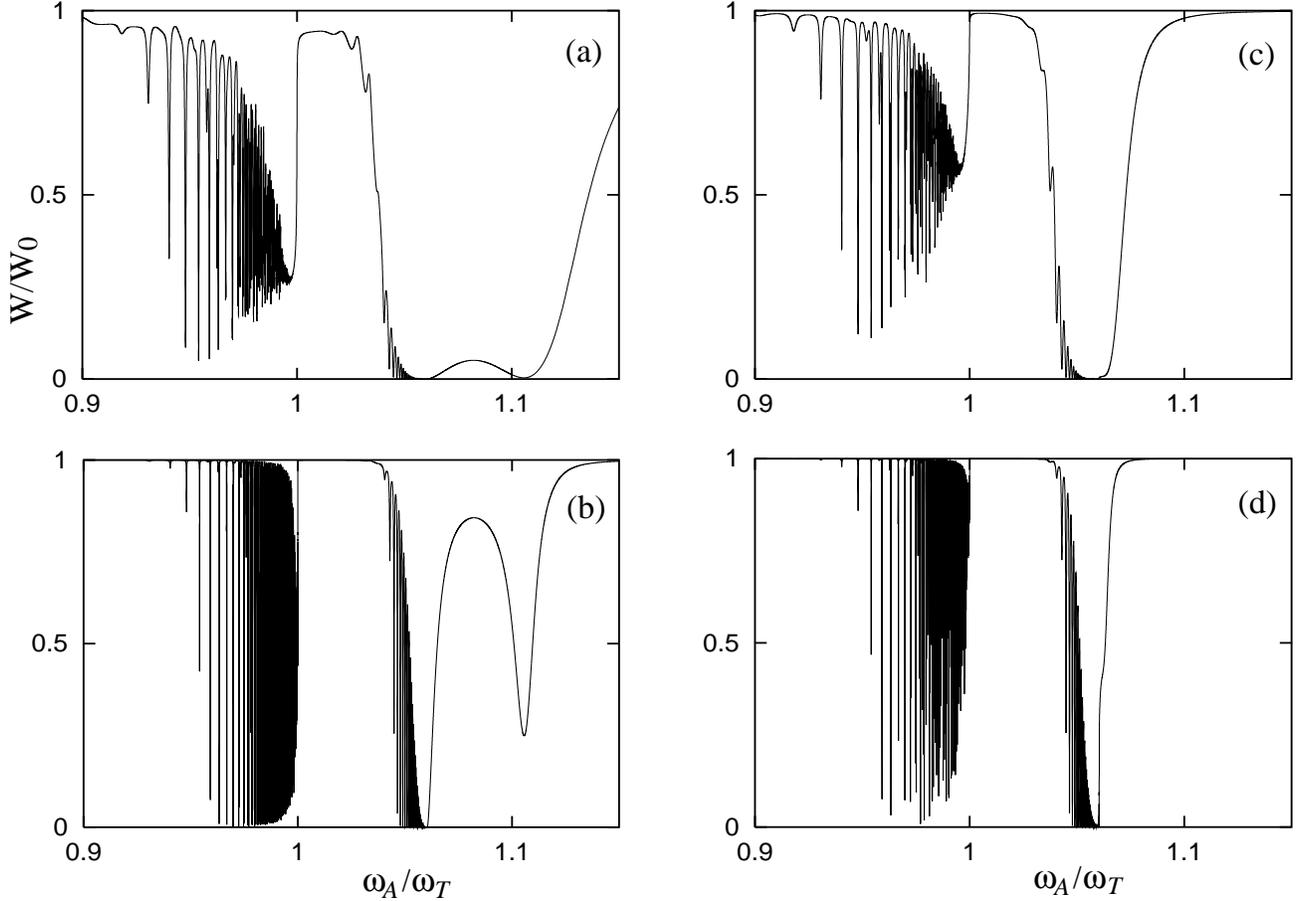,width=1\linewidth}
\end{center}
\caption{
The fraction of emitted radiation energy, Eq.~\protect(\ref{e34}),  
as a function of the atomic transition frequency for 
$\gamma/\omega_T$ and $\Delta r/\lambda_T$ equal to
(a) $10^{-4}$ and $0.02$, 
(b) $10^{-6}$ and $0.02$, 
(c) $10^{-4}$ and $0.1$, and
(d) $10^{-6}$ and $0.1$, respectively.
The other parameters are the same as in  Fig.~\protect\ref{f6}.
}
\label{f8}
\vspace*{5mm}
\end{figure}
\begin{multicols}{2}
\begin{eqnarray}
\label{e34}
\lefteqn{
      {W\over W_0} = {3A_0\over 2 A^\perp}
      \sum_{l=1}^\infty l(l+1)(2l+1)
}
\nonumber\\&&\hspace{2ex} \times\, 
     {1\over ({k_Ar_A})^2}
     \left|j_l({k_Ar_A}) + {\cal B}_l^N\!(\omega_A)
            h_l^{(1)}({k_Ar_A}) \right|^2.
\end{eqnarray}
Recall that \mbox{$W/W_0$ $\!=$ $\!1$} implies fully radiative decay, 
while \mbox{$W/W_0$ $\!=$ $\!0$} fully nonradiative one. 

The dependence of the ratio $W/W_0$ on the atomic
transition frequency is illustrated in Fig.~\ref{f8}.
The minima at the WG field resonance frequencies
indicate that the nonradiative decay is enhanced relative
to the radiative one. Obviously,
photons at these frequencies are captured inside the
microsphere for some time, and hence the probability
of photon absorption is increased. 
For transition frequencies inside the band gap, two regions
can be distinguished.
In the low-frequency region where low-order 
SG waves are typically excited radiative decay dominates. Here,
the light penetration depth into the sphere is small and
the probability of a photon being absorbed is small as well.  
With increasing atomic transition frequency
the penetration depth increases and the chance of 
a photon to escape drastically diminishes. As a result,
nonradiative decay dominates. Clearly, the strength of the
effect decreases with decreasing material absorption  
[compare Figs.~\ref{f8}(a) and (c) with Figs.~\ref{f8}(b)
and (d) respectively]. 

{F}rom Figs.~\ref{f8}(a) and (b) two well
pronounced minima of the
totally emitted light energy,
i.e., noticeable maxima of the energy transfer to the matter,  
are seen for transition
frequencies inside the band gap.
The first minimum results from the 
overlapping high-order SG waves that mainly underly 
absorption losses. The second one
is observed at the longitudinal resonance frequency
of the medium.
It can be attributed to the atomic near-field interaction 
with the longitudinal component of the medium-assisted 
electromagnetic field, the strength of the longitudinal 
field resonance being proportional to $\epsilon_I$.
Hence,
the dip at the longitudinal frequency
of the emitted radiation energy reduces with decreasing
material absorption [compare Fig.~\ref{f8}(a) with 
Fig.~\ref{f8}(b)], and it disappears when 
the atom is moved sufficiently away from the surface
[compare Figs.~\ref{f8}(a) and (b) with Figs.~\ref{f8}(c)
and (d) respectively].
%
\begin{figure}[!t!]
\noindent
\begin{center}
\epsfig{figure=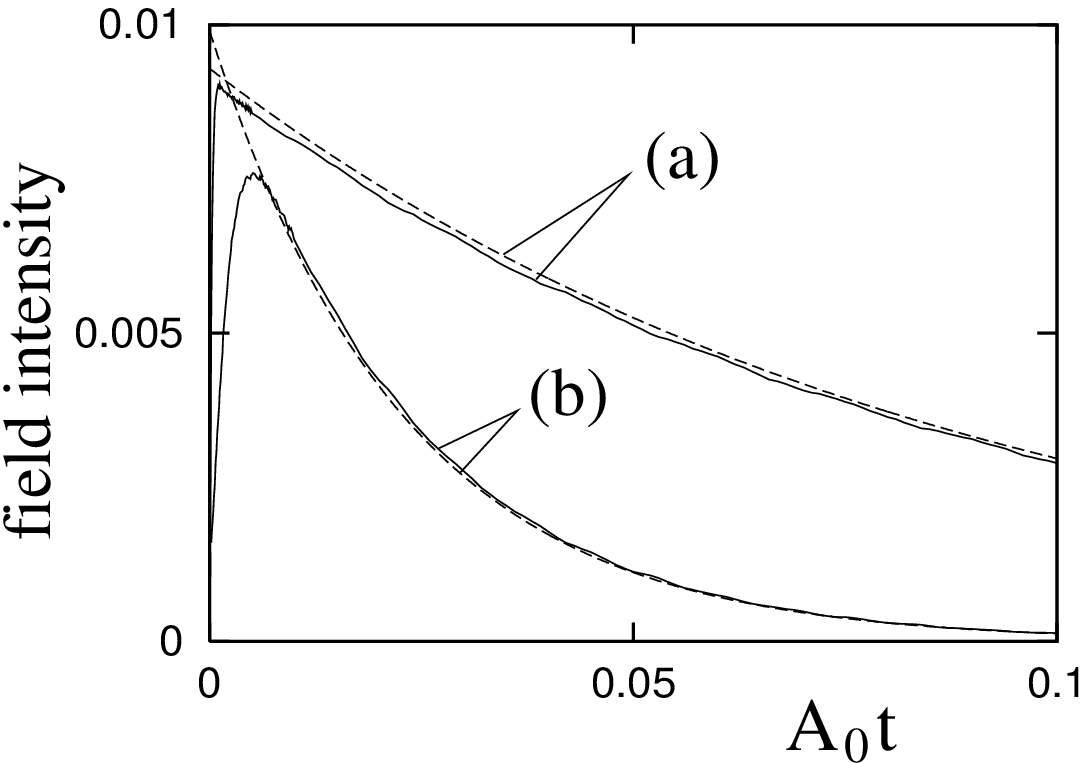,width=1\linewidth}
\end{center}
\caption{
Exact [Eq.~(\protect\ref{e14a}), solid lines] and
approximate [Eq.~(\protect\ref{e15}), dashed lines]
time evolution of the far-field intensity 
$I({\bf r},t)/(k_A^3\mu/4\pi\epsilon_0)^2$ at a fixed
point of observation for a radially oscillating
transition dipole moment (\mbox{$A_0/\omega_T$ $\!=$ $\!10^{-7}$}, 
\mbox{$R$ $\!=$ $\!2\,\lambda_T$}, \mbox{$\Delta r$ $\!=$
$\!0.02\,\lambda_T$}, \mbox{$r$ $\!=$ $\!20\,\lambda_T$},
\mbox{$\theta$ $\!=$ $\!3$}, 
\mbox{$\omega_P$ $\!=$ $\!0.5\,\omega_T$},  
\mbox{$\gamma$ $\!=$ $\!10^{-4}\omega_T$}).
(a) \mbox{$\omega_A$ $\!=$ $\!0.91779\,\omega_T$}  
(TM$_{14,1}$ WG wave with \mbox{$Q$ $\!\sim$ $\!10^3$}). 
(b) \mbox{$\omega_A$ $\!=$ $\!0.94042\,\omega_T$} 
(TM$_{16,1}$ WG wave with \mbox{$Q$ $\!\sim$ $\!10^4$}). 
}
\label{f10}
\end{figure}
%

\subsubsection{Temporal evolution}
\label{sec3.4.3}

Throughout this paper we have restricted our attention to the
weak-coupling regime where the excited atomic state decays
exponentially, Eq.~(\ref{e14b}). When retardation is disregarded,
then the intensity of the emitted light (at some chosen space point)
simply decreases exponentially, Eq.~(\ref{e15}).
To study the effect of retardation, we have also performed
the frequency integral in the exact equation (\ref{e14a})
numerically.

Typical examples of the temporal evolution of the far-field
intensity are shown in Fig.~\ref{f10} for a radially oriented
transition dipole moment in the case when the atomic transition
frequency coincides with the frequency of a WG wave.
Whereas the long-time behavior of the intensity of the
emitted light is, with little error, exponential,
the short-time behavior (on a time scale given by the
atomic decay time) sensitively depends on the quality factor.
The observed delay between the upper-state atomic population
and the intensity of the emitted light can be quite
large for a high-$Q$ microsphere, because the time that
a photon spends in the sphere increases with the $Q$\,value.  
Further, in the short-time domain some kink-like
fine structure is observed, which obviously reflects the
different arrival times associated with multiple reflections. 

\begin{figure}[!t!]
\noindent
\begin{center}
\epsfig{figure=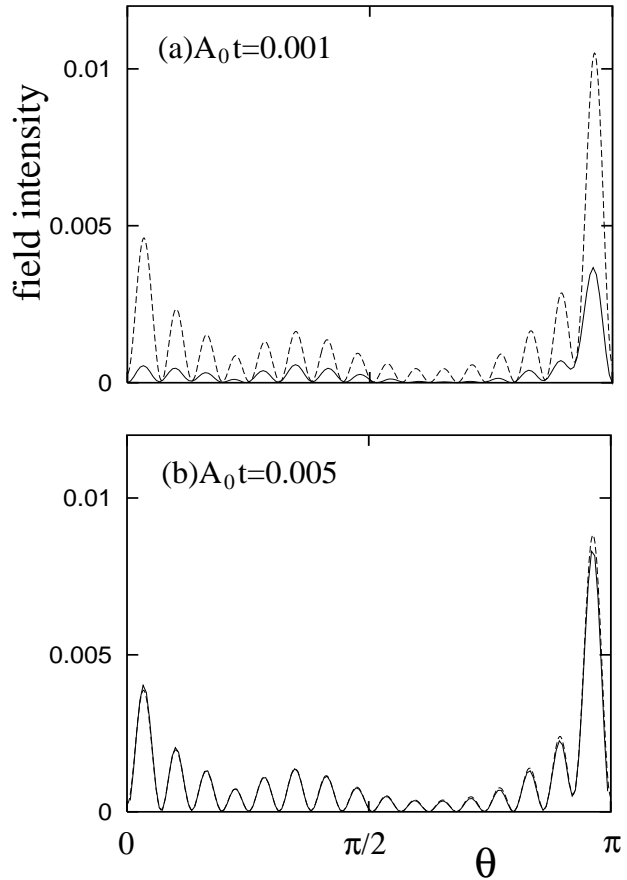,width=1\linewidth}
\end{center}
\caption{
Exact [Eq.~(\protect\ref{e14a}), solid lines] and
approximate [Eq.~(\protect\ref{e15}), dashed lines]
angular distribution of the  far-field intensity 
$I({\bf r},t)/(k_A^3\mu/4\pi\epsilon_0)^2$
at different times for a radially oscillating
transition dipole moment [for the parameters, see
Fig.~\protect\ref{f10}, curves (b)].
}
\label{f11}
\end{figure}

The results in Fig.~\ref{f10} refer to a fixed
space point. Figure~\ref{f11} illustrates the transient
behavior of the spatial distribution of the emitted
light. 
In particular, it is seen that some time is necessary
to build up the sphere-assisted spatial distribution of the
emitted light which is typically observed for longer times
when the approximation (\ref{e15}) applies.
%
%


\section{Metallic microsphere}
\label{sec4}

\begin{figure}[!t!]
\noindent
\begin{center}
\epsfig{figure=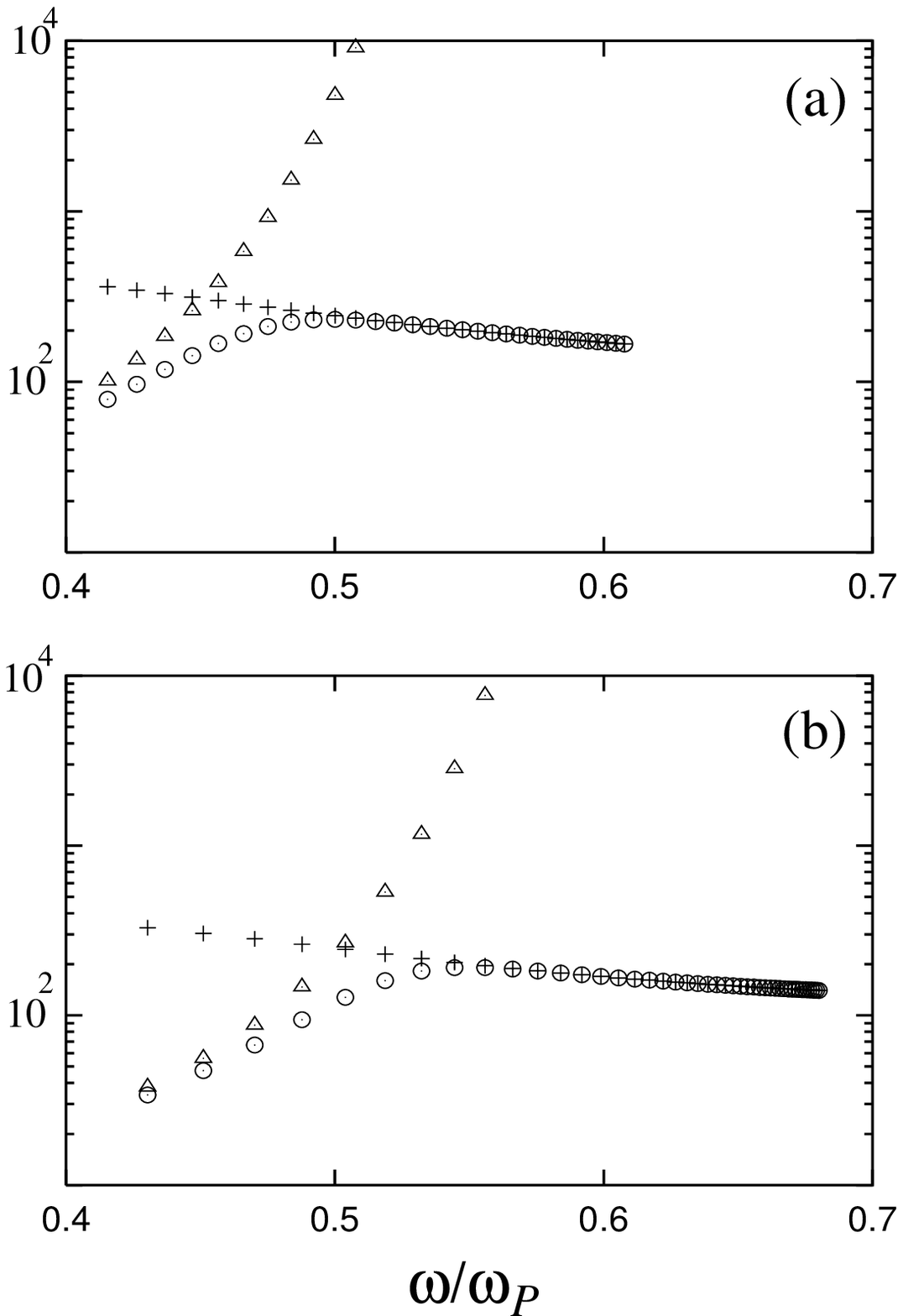,width=1.\linewidth}
\end{center}
\caption{
Quality factors $Q_{rad}$ ($\triangle$),
$Q_{abs}$ (+), and $Q_{tot}$ ({\large $\circ$})
of TM$_l$ SG field resonances in 
a metallic microsphere of complex 
$\epsilon(\omega)$ [Eq.~(\protect\ref{e10}),
\mbox{$\omega_T$ $\!=$ $\!0$},
\mbox{$\gamma/\omega_P$ $\!=$ $\!0.005$}].
\mbox{(a) $R$ $\!=$ $\!10\,\lambda_P$},
\mbox{$30$ $\!\leq$ $\!l$ $\!\leq$ $\!60$}.
\mbox{(b) $R$ $\!=$ $\!5\,\lambda_P$},
\mbox{$16$ $\!\leq$ $\!l$ $\!\leq$ $\!60$}.
All resonances obey, in accordance with the
condition (\protect\ref{e12}), the relation
\mbox{$\omega/\omega_P$ $\!<$ $\!1/\sqrt{2}$ $\!\simeq$
$\!0.71$}.
}
\label{f14}
\vspace*{5mm}
\end{figure}

\begin{figure}[!t!]
\noindent
\begin{center}
\epsfig{figure=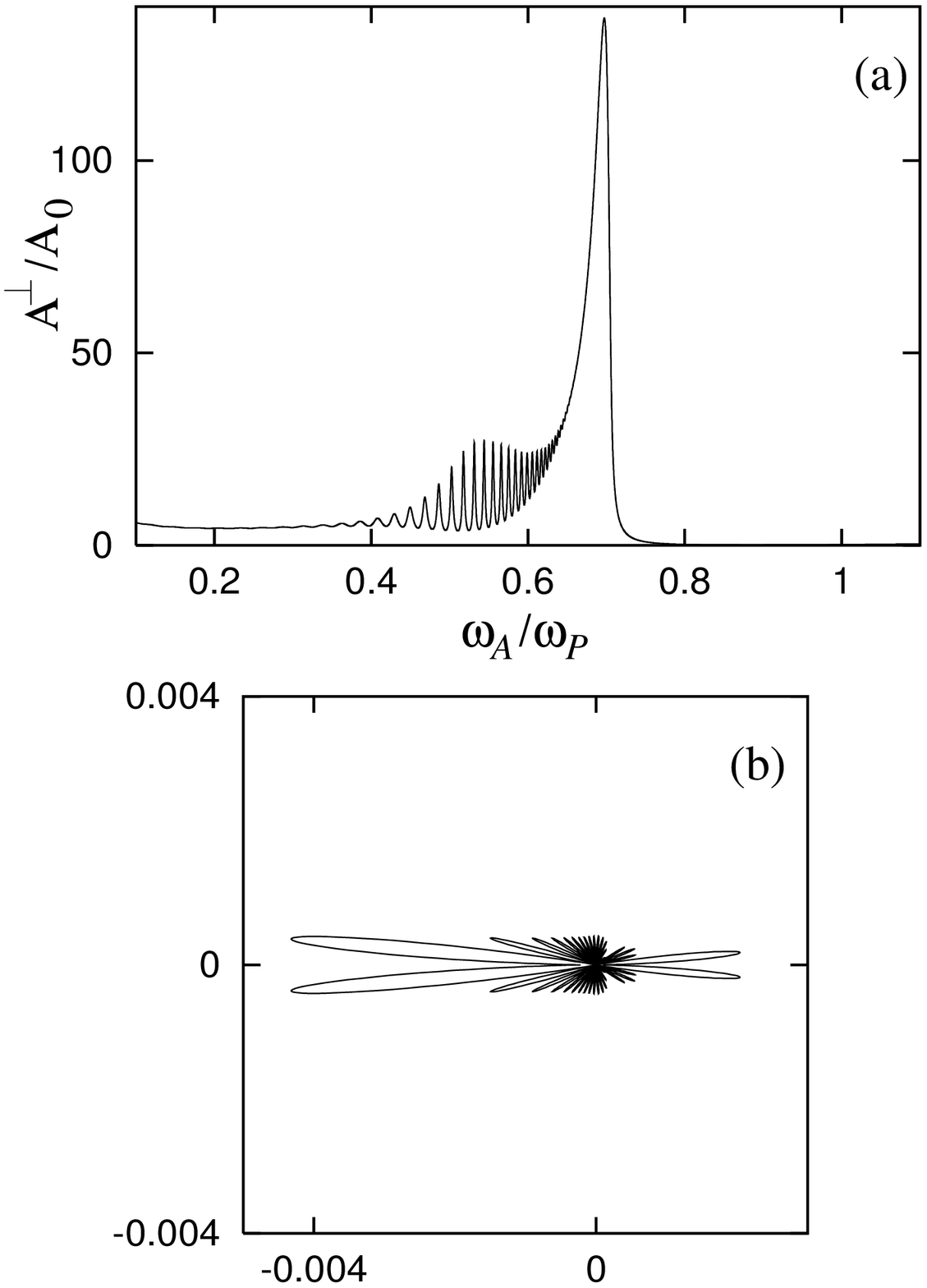,width=1.\linewidth}
\end{center}
\caption{
(a) Decay rate of an atom near a a metallic microsphere of complex 
permittivity $\epsilon(\omega)$ [Eq.~(\protect\ref{e10}),
\mbox{$\omega_T$ $\!=$ $\!0$},
\mbox{$\gamma/\omega_P$ $\!=$ $\!0.005$}]
as a function of the transition frequency for a radially oriented
transition dipole moment (\mbox{$R$ $\!=$ $\!5\lambda_P$},
\mbox{$\Delta r$ $\!=$ $\!0.1\,\lambda_P$}). 
(b) Polar diagram of the normalized far-field emission pattern 
$|{\bf F}^\perp({\bf r},{\bf r}_A,\omega_A)|^2/
(k_A^3\mu / 4\pi \epsilon_0)^2$
for \mbox{$r$ $\!=$ $\!50\,\lambda_P$} and   
\mbox{$\omega_A/\omega_P$ $\!=$ $\!0.5026$}, the   
other parameters being the same as in (a).
}
\label{f15}
\vspace*{5mm}
\end{figure}

Setting in Eq.~(\ref{e10}) \mbox{$\omega_T$ $\!=$ $\!0$},
we obtain (within the Drude-Lorentz model) the permittivity
of a metal. Hence, the results derived for the
band gap of a dielectric microsphere also applies,
for appropriately chosen values of $\omega_P$ and $\gamma$,
to a metallic sphere. The examples plotted in
Figs.~\ref{f14} and \ref{f15} refer to silver 
(\mbox{$\omega_P$ $\!=$ $\!1.32\times 10^{16}$\,s$^{-1}$}
and \mbox{$1/\gamma$ $\!=$ $\!1.45\times 10^{-14}$\,s}
\cite{17b}).

Figure \ref{f14} shows the positions and quality factors
of SG field resonances for different microsphere radii.
The behavior is quite similar to that observed for
a dielectric microsphere [cf. Fig.~\ref{f2}].
The radiative losses decrease with increasing radius of the
sphere, while the losses due to material absorption are
less sensitive to the microsphere size [compare
Figs.~\ref{f14}(a) and (b)].
It is further seen that $Q_{rad}$ increases with the angular 
momentum number of the SG wave, while $Q_{abs}$ slightly
decreases. 
%
%
As a result, the radiative losses again dominate for 
low-order resonances and the absorption losses for 
high-order ones. Since absorption tends to be larger 
in metals than in dielectrics, the dominance of material
absorption can already set in at lower-order resonances.
Note that even for metals
the relationship (\ref{e10b1})
typically still holds, so that
Eqs.~(\ref{e11}) -- (\ref{e12.1}) apply.

The dependence of the decay rate on the transition
frequency of an excited atom placed near a metallic
microsphere is illustrated in Fig.~\ref{f15}(a) for
a radially oriented transition dipole moment. An example
of the emission pattern for the case when the atomic
transition frequency coincides with the frequency of a
SG wave is shown in Fig.~\ref{f15}(b).
When the radius of the microsphere becomes too small,
then SG waves cannot be excited. In particular,
in \cite{17c} it was assumed that $R$ $\!\ll$ $\!\lambda_P$,
and thus the resonances shown in Figs.~\ref{f14} and \ref{f15}(a)
could not be found. It is worth noting that,
in contrast to dielectric matter, a large absorption 
in metals can substantially
%
%
\noindent
 enhance the near-surface
divergence of the decay rate, [Eqs.~(\ref{E21}) and (\ref{E22})],
which is in agreement with experimental observations 
of the fluorescence from a thin layer of optically 
excited organic-dye molecules that were separated from 
a planar metal surface by a dielectric layer of known 
thickness \cite{28}.


\section{Conclusions}
\label{sec5}

We have applied the recently developed formalism \cite{12} 
to the problem of spontaneous decay of an excited atom
near a dispersing and absorbing microsphere. Basing
the calculations on a complex permittivity of
Drude-Lorentz type, which satisfies the Kramers-Kronig
relations, we have been able to study the dependence
of the decay rate on the transition frequency 
for arbitrary frequencies. We have shown that
the decay can be substantially enhanced when
the transition frequency is tuned to either a WG field
resonance below the band gap (for a dielectric sphere)
or a SG field resonance inside the band gap (for a dielectric
sphere or a metallic sphere).

Whereas for both low-order WG field resonances and low-order
SG field resonances radiative losses dominate, high-order
field resonances mainly suffer from material absorption. Accordingly,
spontaneous decay changes from being mainly radiative
to being mainly nonradiative, when the transition
frequency (tuned to a field resonance) in the respective
frequency interval increases.
We have further
shown that in the presence of strong material absorption 
the decay rate drastically raises as the atom approaches the 
surface of the microsphere, because of near-field assisted
energy transfer from the atom to the medium. Thus, the effect
is typically observed for metals.

When radiative losses dominate, the emission pattern is
highly structured, and a substantial fraction of the light
is emitted backward and forward within small polar angles
with respect to the tie line between the atom and the
center of the sphere. With increasing absorption
this directional characteristic is lost. 
When absorption losses dominate,
the (weakened) emission pattern takes a form that
is typical of reflection at a mirror. 

In the paper we have restricted our attention to the weak-coupling
regime, assuming that the excited atomic state decays
exponentially. Obviously, when the
atomic transition frequency coincides with a resonance
frequency of the cavity-assisted field, the strong-coupling
regime may be realized. In particular, SG waves seem to be
best suited for that regime, because of the noticeable
enhancement of spontaneous emission. The calculations
could be performed in a similar way as in Ref.~\cite{12}.

\acknowledgements

We thank F. Lederer, S. Scheel, and E. Schmidt for fruitful 
discussions. H.T.D. is grateful to the Alexander von 
Humboldt Stiftung and the Vietnamese Basic Research
Program for financial support. This work was supported 
by the Deutsche Forschungsgemeinschaft.


\appendix

\section{The Green tensor}
\label{appA}

The Green tensor of a microsphere of radius $R$ (reg\-\mbox{ion 2)}
embedded in vacuum (region 1) can be decomposed into
two parts \cite{26}, 
\begin{equation} 
\label{A1}
\bbox{G}({\bf r},{\bf r'},\omega) 
= \bbox{G}^{(s)}({\bf r},{\bf r'},\omega) \delta_{fs} 
+ \bbox{G}^{(fs)}({\bf r},{\bf r'},\omega),
\end{equation}
where $\bbox{G}^{(s)}({\bf r},{\bf r'},\omega)$ represents 
the contribution of the direct waves from the source
in an unbounded space,
and $\bbox{G}^{(fs)}({\bf r},{\bf r'},\omega)$ is the
scattering part that describes the contribution of the multiple
reflection \mbox{($f$ $\!=$ $\!s$)} and transmission
\mbox{($f$ $\!\neq$ $\!s$)} waves ($f$ and $s$, respectively,
refer to the regions where are the field and source points
${\bf r}$ and ${\bf r}'$).
In particular, 
$\bbox{G}^{(s)}$, $\bbox{G}^{(11)}$, and
$\bbox{G}^{(22)}$ can be given by \cite{26}
\begin{eqnarray}
\label{A2}
\lefteqn{
       \bbox{G}^{(s)}({\bf r},{\bf r}',\omega)
       = {{\bf e}_r{\bf e}_r\over k_s^2} \delta(r-r')
}
\nonumber\\&&\hspace{2ex}
       +{ik_s\over 4\pi} \sum_{e\atop o} 
       \sum_{l=1}^\infty \sum_{m=0}^l 
       {2l+1\over l(l+1)} {(l-m)!\over(l+m)!}\,(2\!-\!\delta_{0m})
\nonumber\\&&\hspace{2ex}
       \times
       \left[ {\bf M}^{(1)}_{{e \atop o}lm} ({\bf r},k_s)
       {\bf M}_{{e \atop o}lm} ({\bf r}',k_s) \right.
\nonumber\\&&\hspace{4ex}       
       \left. +\,{\bf N}^{(1)}_{{e \atop o}lm} ({\bf r},k_s)
       {\bf N}_{{e \atop o}lm} ({\bf r}',k_s)
        \right]
\end{eqnarray}
if \mbox{$r$ $\!\ge$ $\!r'$}, and
\mbox{$\bbox{G}^{(s)}({\bf r},{\bf r}',\omega)$  $\!=$
$\!\bbox{G}^{(s)}({\bf r}',{\bf r},\omega)$}
if \mbox{$r$ $\!<$ $\!r'$},
\begin{eqnarray}
\label{A3}
\lefteqn{
       \bbox{G}^{(11)}({\bf r},{\bf r'},\omega)
}
\nonumber\\&&\hspace{2ex}
       = {ik_1\over 4\pi} \sum_{e\atop o} 
       \sum_{l=1}^\infty \sum_{m=0}^l 
       {2l+1\over l(l+1)} {(l-m)!\over(l+m)!}\,(2\!-\!\delta_{0m})
\nonumber\\&&\hspace{2ex}\times 
       \Bigl[ {\cal B}^M_l (\omega)
       {\bf M}^{(1)}_{{e \atop o}lm} ({\bf r},k_1)
       {\bf M}^{(1)}_{{e \atop o}lm} ({\bf r}',k_1)
\nonumber\\&&\hspace{2ex}
       +\; {\cal B}^N_l (\omega)
       {\bf N}^{(1)}_{{e \atop o}lm} ({\bf r},k_1)
       {\bf N}^{(1)}_{{e \atop o}lm} ({\bf r}',k_1) \Bigr]  
\label{A3a}
\end{eqnarray}
\mbox{($r,r'$ $\!>$ $\!R$)},
\begin{eqnarray}
\lefteqn{
       \bbox{G}^{(22)}({\bf r},{\bf r'},\omega)
}
\nonumber\\&&\hspace{2ex}
       ={ik_2\over 4\pi} \sum_{e\atop o} 
       \sum_{l=1}^\infty \sum_{m=0}^l 
       {2l+1\over l(l+1)} {(l-m)!\over(l+m)!}\,(2\!-\!\delta_{0m})
\nonumber\\&&\hspace{2ex}\times 
       \Bigl[ {\cal C}^M_l (\omega)
       {\bf M}_{{e \atop o}lm} ({\bf r},k_2)
       {\bf M}_{{e \atop o}lm} ({\bf r}',k_2)
\nonumber\\&&\hspace{2ex}
       +\; {\cal C}^N_l (\omega)
       {\bf N}_{{e \atop o}lm} ({\bf r},k_2)
       {\bf N}_{{e \atop o}lm} ({\bf r}',k_2) \Bigr]    
\end{eqnarray}
\mbox{($r,r'$ $\!<$ $\!R$)}, where
\begin{equation}
\label{A3b}
       k_1={\omega\over c}, \qquad 
       k_2=\sqrt{\epsilon(\omega)}\,{\omega\over c}\,. 
\end{equation}
${\bf M}$ and ${\bf N}$ represent TE and TM waves, respectively,
\begin{eqnarray}
\label{A4}
\lefteqn{
       {\bf M}_{{e \atop o}lm}({\bf r},k) 
       = \mp {m \over \sin\theta} j_l(kr)
       P_l^m(\cos\theta) {\sin\choose\cos} (m\phi) {\bf e}_{\theta}
}
\nonumber\\&&\hspace{12ex}
       -\, j_l(kr) \frac{dP_l^m(\cos\theta)}{d\theta} 
       {\cos\choose\sin} (m\phi) {\bf e}_{\phi} \,,
\end{eqnarray}
\begin{eqnarray}
\label{A5}
\lefteqn{
       {\bf N}_{{e \atop o}lm}({\bf r},k) 
       = {l(l\!+\!1)\over kr} j_l(kr) 
       P_l^m(\cos\theta) {\cos\choose\sin} (m\phi) {\bf e}_r
}
\nonumber\\&&\hspace{12ex}
       +\,{1\over kr} \frac{d[rj_l(kr)]}{dr} 
       \Biggl[
       \frac{dP_l^m(\cos\theta)}{d\theta} 
       {\cos\choose\sin} (m\phi) {\bf e}_{\theta}
\nonumber \\&&\hspace{12ex}
       \mp\, {m \over \sin\theta}
       P_l^m(\cos\theta) {\sin\choose\cos} (m\phi) {\bf e}_{\phi}
       \Biggr], 
\end{eqnarray}
with $j_l(x)$ and $P_l^m(x)$
being respectively the spherical Bessel function of the first kind
and
the associated Legendre function. 
The superscript ${(1)}$ in Eqs.~(\ref{A2}) and (\ref{A3}) indicates that 
in Eqs.~(\ref{A4}) and (\ref{A5}) the spherical Bessel function $j_l(x)$ 
has to be replaced by the first-type spherical Hankel function $h^{(1)}_l(x)$.
The coefficients ${\cal B}^{M,N}_l$ and ${\cal C}^{M,N}_l$
in Eqs.~(\ref{A3}) and (\ref{A3a}) are defined by
\begin{eqnarray}
\label{A6}
\lefteqn{
       {\cal B}^M_l(\omega) 
}
\nonumber\\&&\hspace{2ex}
       = - \frac 
       {\bigl[ z_2j_l(z_2)\bigr]' j_l(z_1)
       - \bigl[ z_1j_l(z_1)\bigr]' j_l(z_2) }
       {\bigl[ z_2j_l(z_2)\bigr]' h_l^{(1)}(z_1)
       -  j_l(z_2) \bigl[z_1 h_l^{(1)}(z_1)\bigr]' }\,,
\end{eqnarray}
\begin{eqnarray}
\label{A7}
\lefteqn{
       {\cal B}^N_l(\omega) 
}
\nonumber\\&&\hspace{2ex}
       = -  \frac 
       { \epsilon(\omega) 
       j_l(z_2) \bigl[z_1 j_l(z_1)\bigr]'
       - j_l(z_1) \bigl[z_2 j_l(z_2)\bigr]' }
       { \epsilon(\omega) 
       j_l(z_2) \bigl[ z_1 h_l^{(1)}(z_1)\bigr]' 
       - \bigl[z_2 j_l(z_2)\bigr]' h_l^{(1)}(z_1)} \,,
\end{eqnarray}
\begin{eqnarray}
\label{A7a}
\lefteqn{
       {\cal C}^M_l(\omega)  
}
\nonumber\\&&\hspace{1ex}
       = - \frac 
       {\bigl[ z_2h_l^{(1)}(z_2)\bigr]' h_l^{(1)}(z_1)
       - \bigl[ z_1h_l^{(1)}(z_1)\bigr]' h_l^{(1)}(z_2) }
       {\bigl[ z_2j_l(z_2)\bigr]' h_l^{(1)}(z_1)
       -  j_l(z_2) \bigl[z_1 h_l^{(1)}(z_1)\bigr]' }\,,
\end{eqnarray}
\begin{eqnarray}
\lefteqn{
       {\cal C}^N_l(\omega)  
}
\nonumber\\&&\hspace{0ex}
       = - \frac 
       { \epsilon(\omega) 
       h_l^{(1)}(z_2) \bigl[z_1 h_l^{(1)}(z_1)\bigr]'
       - h_l^{(1)}(z_1) \bigl[z_2 h_l^{(1)}(z_2)\bigr]' }
       { \epsilon(\omega) 
       j_l(z_2) \bigl[ z_1 h_l^{(1)}(z_1)\bigr]' 
       - \bigl[z_2 j_l(z_2)\bigr]' h_l^{(1)}(z_1)}\,, 
\nonumber\\&&\hspace{0ex}
\label{A7b}
\end{eqnarray}
where
\begin{equation}
\label{A8}
       z_i = k_iR \ .
\end{equation}
%
Note that the relations 
\begin{equation} 
\int_{-1}^1 dx\,P^n_l(x)P^n_m(x) =
\frac{(l+n)!}{(l-n)!(l+1/2)}\,\delta_{lm}
\label{A10}
\end{equation}
and 
\begin{equation} 
h^{(1)}_l(z) \rightarrow
z^{-1}\exp\!\left[i(z\!-\!l\pi/2\!-\!\pi/4)\right]
\quad{\rm if}\quad 
|z| \rightarrow \infty
\label{A11}
\end{equation}
are valid \cite{18}.


\section{SG field resonances}
\label{appB}

For large value of \mbox{$\nu$ $\!=$ $\!l+1/2$} the following
asymptotic expansions are valid \cite{18}: 
\begin{equation}
\label{B1}
      J_\nu(nx) \sim 
      {\exp[\nu(\tanh\alpha\!-\!\alpha)] \over 
       \sqrt{2\pi\nu\tanh\alpha}}
      \left[ 1\!+\! \sum_{k=1}^\infty
      {u_k(\coth\alpha)\over\nu^k}\right]\! ,
\end{equation}
\begin{eqnarray}
\label{B2}
\lefteqn{
      Y_\nu(x) \sim 
      -{\exp[\nu(\beta\! -\!\tanh\beta)] \over 
       \sqrt{{1\over 2}\pi\nu\tanh\beta}}
}
\nonumber\\&&\hspace{10ex}\times\,
       \left[ 1+ \sum_{k=1}^\infty (-1)^k
      {u_k(\coth\beta)\over\nu^k}\right]\!,
\end{eqnarray}
\begin{eqnarray}
\label{B3}
\lefteqn{
      J'_\nu(nx) \sim
      \sqrt{{\sinh 2\alpha\over 4\pi\nu}} \,
      \exp[\nu(\tanh\alpha-\alpha)] 
}
\nonumber\\&&\hspace{10ex}\times\,
      \left[ 1+ \sum_{k=1}^\infty
      {v_k(\coth\alpha)\over\nu^k}\right]\!,
\end{eqnarray}
\begin{eqnarray}
\label{B4}
\lefteqn{
      Y'_\nu(x) \sim 
      \sqrt{{\sinh 2\beta\over \pi\nu}} 
      e^{\nu(\beta -\tanh\beta)} 
}
\nonumber\\&&\hspace{10ex}\times\,
\left[ 1+ \sum_{k=1}^\infty (-1)^k
      {v_k(\coth\beta)\over\nu^k}\right]\!,
\end{eqnarray}
[$Y_\nu(x)$ - Neumann function], where
\begin{eqnarray}
\label{B5}
&\displaystyle 
       x={\omega\over c}R,
\\&\displaystyle
\label{B6}
       \cosh\alpha = {\nu\over nx}, \quad 
       \cosh\beta = {\nu\over x} \,,
\end{eqnarray}
and $u_k$ and $v_k$ are given in Ref.~\cite{18}.
To find the
leading terms in Eqs.~(\ref{e3}) and (\ref{e4}),
we thus may write 
\begin{equation}
\label{B7}
     {J'_\nu(nx) \over J_\nu(nx) } \sim |\sinh \alpha|,
\end{equation}
and
\begin{equation}
\label{B8}
     {H'_\nu(x) \over H_\nu(x) } \sim 
     {Y'_\nu(x) \over Y_\nu(x) } \sim - |\sinh\beta| 
\end{equation}
to obtain, 
\begin{equation}
\label{B9}
\sqrt{\nu^2-x^2}+\sqrt{\nu^2-\epsilon x^2} = 0 
\end{equation}
for TE waves, and
\begin{equation}
\label{B10}
\epsilon\sqrt{\nu^2-x^2}+\sqrt{\nu^2-\epsilon x^2} = 0
\end{equation}
for TM waves. Here we have used relationships (\ref{B6}),
and we have assumed that $x$ scales as $\nu$ to discard the last term 
in Eq. (\ref{e4}).

Obviously, Eq. (\ref{B9}) for TE waves 
has no solution, except for the trivial case of \mbox{$\epsilon$
$\!=$ $1$}. Equation (\ref{B10}) for TM waves can be rewritten as
\begin{equation}
\label{B11}
       x=\nu\sqrt{1+\epsilon^{-1}}\,, 
\end{equation}
which just implies condition (\ref{e11a}). 
The higher-order corrections can be obtained by writing
\begin{equation}
\label{B12}
       x=\nu\sqrt{1+\epsilon^{-1}} \left[1+
       \sum_{k=1}^\infty {c_k \over  \nu^{k}}\right], 
\end{equation}
expanding all the quantities in Eq.~(\ref{e4}) in powers
of $\nu^{-1}$ and identifying the corresponding terms.


\section{Near-surface limit}
\label{appC}

Using the asymptotic Bessel-function expansion \cite{18}
\begin{eqnarray}
\label{C1}
      J_\nu(z) \sim {1\over \sqrt{2\pi\nu}} 
      \left(ez\over 2\nu\right)^\nu\!,
\quad
      Y_\nu(z) \sim -{2\over \sqrt{\pi\nu}} 
      \left(ez\over 2\nu\right)^{-\nu}\!,
\end{eqnarray}
($|\nu|$ $\!\gg$ $\!1$), the coefficient ${\cal B}^N_l(\omega_A)$,
Eq.~(\ref{A7}), can be given in the form of
\begin{eqnarray}
\label{C3}
\lefteqn{
      {\cal B}^N_l(\omega_A)
      \left[ h^{(1)}_l({k_Ar_A}) \over {k_Ar_A} \right]^2
}
\nonumber\\&&\hspace{2ex}
      \sim {1\over i ({k_Ar_A})^3 (2l+1)}
      {\epsilon(\omega_A)-1 \over \epsilon(\omega_A)+1}
      \left(R\over r_A\right)^{2l+1} .
\end{eqnarray}
When the atom is located very close to the surface of the
microsphere, i.e.,  \mbox{$r_A$ $\!\gtrsim$ $\!R$},
then from Eq.~(\ref{C3}) it follows that 
the series in Eq.~(\ref{e17}) converges very slowly.
Hence,
it is a good approximation to apply equation (\ref{C3}) to the 
terms with small $l$ as well. In this way, we derive  
\begin{eqnarray}
\label{C4}
\lefteqn{
      \sum_{l=1}^\infty l(l+1)(2l+1)
      {\cal B}^N_l(\omega_A)
      \left[ h^{(1)}_l({k_Ar_A}) \over {k_Ar_A} \right]^2
}
\nonumber\\&&\hspace{15ex}
      \sim \,{1\over 4i} 
      {\epsilon(\omega_A)-1 \over \epsilon(\omega_A)+1}
      {1\over (\Delta r)^3}\,.
\end{eqnarray}
Substitution of this expression into Eq.~(\ref{e17}) then
yields the leading term in Eq.~(\ref{E21}). The leading term in
Eq.~(\ref{E22}) can be derived in a similar fashion.


\end{multicols}
\end{document}